\documentclass[manuscript,screen]{acmart}
\copyrightyear{2026}
\acmYear{2026}

\AtBeginDocument{%
  }

\setcopyright{none}

\acmDOI{XXXXXXX.XXXXXXX}

\usepackage{multirow}
\usepackage{subcaption}
\begin{document}

\title{AnomalyGen: Enhancing Log-Based Anomaly Detection with Code-Guided Data Augmentation}

\author{Xinyu Li}
\email{lixy769@mail2.sysu.edu.cn}
\affiliation{%
  \institution{Sun Yat-sen University}
  \city{Zhuhai}
  \country{China}
}

\author{Yintong Huo}
\email{ythuo@smu.edu.sg}
\affiliation{%
  \institution{Singapore Management University}
  \country{Singapore}
  \city{Singapore}
}

\author{Chenxi Mao}
\email{maochx5@mail2.sysu.edu.cn}
\affiliation{%
  \institution{Sun Yat-sen University}
  \city{Zhuhai}
  \country{China}
}

\author{Shiwen Shan}
\email{shanshw@mail2.sysu.edu.cn}
\affiliation{%
  \institution{Sun Yat-sen University}
  \city{Zhuhai}
  \country{China}
}

\author{Yuxin Su}
\email{suyx35@mail.sysu.edu.cn}
\authornote{Corresponding author.}
\affiliation{%
  \institution{Sun Yat-sen University}
  \city{Zhuhai}
  \country{China}
}

\author{Yanlin Wang}
\email{wangylin36@mail.sysu.edu.cn}
\affiliation{%
  \institution{Sun Yat-sen University}
  \city{Zhuhai}
  \country{China}
}

\author{Zibin Zheng}
\email{zhzibin@mail.sysu.edu.cn}
\affiliation{%
  \institution{Sun Yat-sen University}
  \city{Zhuhai}
  \country{China}
}

\begin{abstract}
Log-based anomaly detection is fundamentally constrained by training data sparsity. 
Our empirical study reveals that public benchmark datasets cover <10\% of source code log templates. 
Consequently, models frequently misclassify unseen but valid execution paths as anomalies, leading to false alarms. 
To address this, we propose AnomalyGen, a novel framework that augments training data by synthesizing labeled log sequences from source code.
AnomalyGen combines log-oriented static analysis with Large Language Model (LLM) reasoning in three stages: 
(1) building Log-Oriented Control Flow Graphs (LCFGs) to enumerate structurally valid execution paths; 
(2) applying LLM Chain-of-Thought (CoT) reasoning to verify logical consistency and generate realistic runtime parameters (e.g., block IDs, IP addresses); and 
(3) labeling generated sequences with domain heuristics.
Evaluations on HDFS and Zookeeper across 12 diverse anomaly detection models show AnomalyGen consistently improves performance. 
Deep learning models achieved average F1-score gains of 2.18\% (HDFS) and 1.69\% (Zookeeper), with an unsupervised Transformer on HDFS jumping from 0.818 to 0.970. 
Ablation results show that both static analysis and LLM-based verification are necessary: removing them reduces F1 by up to 8.7 and 10.7 percentage points, respectively. 
Our framework and datasets are publicly available to facilitate future research.

\end{abstract}

\ccsdesc[500]{Software and its engineering~Automated static analysis}
\ccsdesc[300]{Computing methodologies~Natural language generation}
\ccsdesc[300]{Software and its engineering~Maintaining software}

\keywords{log analysis, anomaly detection, data augmentation, large language models, static analysis}

\maketitle

\section{Introduction}
With the exponential growth in complexity of modern software systems, runtime logs have become the primary data source for debugging and reliability engineering~\cite{zhu2023loghub,mastropaolo2022using,9789917}. 
To mitigate the huge financial losses caused by system failures~\cite{10.1145/3460345}, log-based anomaly detection has emerged as a critical technique for identifying abnormal system behaviors. To this end, current research has developed deep learning models, from LSTMs~\cite{du2017deeplog} to Transformers~\cite{guo2024logformer} and GNNs~\cite{10.1145/3639478.3643084}, to better capture complex log patterns. However, a critical bottleneck arises when deploying these models in practice: their reliance on training data that may not cover all possible log patterns can lead to poor generalization, resulting in some false alarms on unseen workloads. This bottleneck stems from a fundamental data coverage problem. Log-based anomaly detection heavily relies on public benchmarks (e.g., LogHub~\cite{zhu2023loghub}), yet the comprehensiveness of these datasets remains largely unverified. To quantify this gap, we conducted an empirical study comparing log templates collected in public datasets against those defined in system source code. The results, summarized in Table~\ref{tab:coverage_intro}, reveal a fundamental data sparsity problem: existing datasets capture only a microscopic fraction of system behaviors. For instance, the HDFS dataset, one of the most widely used benchmarks, covers only 0.99\% (48 out of 4,846) of the log templates defined in its source code. Coverage for Hadoop, OpenStack, and Zookeeper is similarly low at 1.62\%, 2.86\%, and 8.02\%, respectively. 
As a result, a model trained on such data has never observed most valid execution paths of the system. When those paths appear in production, the model has no basis to recognize them as normal, so it flags them as anomalies. This is not a modeling limitation that better architectures can fix, instead it is a data limitation. No matter how expressive the model, it cannot reliably distinguish between a valid execution path it has never seen and a true anomaly.

\begin{table}[t]
  \centering
  \small
  \caption{A comparison between log templates defined in source code and those observed in public benchmarks. Low coverage indicates existing datasets fail to cover the vast majority of system behaviors.}
  \label{tab:coverage_intro}
  \begin{tabular}{llrrr}
    \toprule
    \textbf{System} & \textbf{Category} & \textbf{\# Source } & \textbf{\# Observed} & \textbf{Coverage} \\
    \midrule
    HDFS            & Distributed Storage & 4,846 & 48  & \textbf{0.99\%} \\
    Hadoop (Common) & Infrastructure      & 15,223 & 248 & \textbf{1.62\%} \\
    OpenStack (Nova)& Cloud Computing     & 2,060 & 59  & \textbf{2.86\%} \\
    Zookeeper       & Coordination        & 998 & 80  & 8.02\% \\
    Spark           & Big Data Processing & 1,749 & 200 & 11.44\% \\
    MapReduce       & Distributed Computing & 1,564 & 200 & 12.79\% \\
    \midrule
    \textbf{Average}& - & \textbf{4,406} & \textbf{131} & \textbf{< 5.0\%} \\
    \bottomrule
  \end{tabular}
\end{table}

Prior work on log data augmentation falls into two categories, each with a fundamental limitation that prevents it from addressing data coverage. 
Transformation-based methods~\cite{10.1145/3338906.3338931} enlarge dataset by randomly inserting, deleting, or shuffling log events within existing sequences. 
Because these operations work purely on observed data, they cannot introduce log templates that were absent from the original dataset and do not expand coverage. 
Static analysis methods~\cite{huo2023autolog} traverse program control flow graphs to enumerate execution paths, which can produce new template sequences. However, they cannot determine realistic values for log parameters (e.g., block IDs, IP addresses), and they struggle to handle data-dependent branches whose outcome depends on runtime variable values. The resulting sequences consist only of template stubs, which limits the performance of detection models that rely on semantic content.

To investigate whether simple data augmentation strategies can improve log-based anomaly detection, we conducted a motivating experiment on the HDFS dataset. We applied two basic augmentation methods: random perturbation (randomly inserting, deleting, or modifying log events within existing sequences) and random resampling (duplicating and recombining existing sequences). Both methods were applied at the same augmentation ratio of 0.01.
As shown in Table~\ref{tab:aug_comparison}, the results reveal a clear challenge: rather than improving detection, these naive approaches consistently degrade performance. Perturbation reduces F1-scores on five out of seven model variants, and resampling does so on five out of seven. For example, the Transformer (semantic) F1-score drops from 0.722 to 0.715 with perturbation and 0.712 with resampling; the LSTM (next\_log, semantic) falls from 0.784 to 0.742 with resampling. Even in cases where perturbation shows marginal gains, the improvements are erratic and disappear entirely under resampling.

These findings highlight a fundamental challenge: effective augmentation requires generating structurally valid and previously unseen execution paths. Simply rearranging or duplicating existing data introduces noise—either by disrupting valid sequence structures or by adding redundant information—rather than closing the coverage gap. This motivates the need for more sophisticated approaches that can synthesize realistic, yet novel, log sequences.

\begin{table}[t]
  \centering
  \small
  \caption{F1-score on HDFS at augmentation ratio 0.01 under two prior augmentation strategies. $\downarrow$ denotes degradation relative to the no-augmentation baseline.}
  \label{tab:aug_comparison}
  \begin{tabular}{lcccc}
    \toprule
    \textbf{Model} & \textbf{No Aug} & \textbf{+Perturbation} & \textbf{+Resample} \\
    \midrule
    Transformer (next\_log)    & 0.862 & \textbf{0.900} & 0.832$\downarrow$ \\
    Transformer (semantic)     & \textbf{0.722} & 0.715$\downarrow$ & 0.712$\downarrow$ \\
    CNN (sequential)           & 0.958 & 0.974 & \textbf{0.980} \\
    CNN (semantic)             & \textbf{0.974} & 0.968$\downarrow$ & 0.969$\downarrow$ \\
    LSTM (sequential)          & \textbf{0.960} & 0.942$\downarrow$ & \textbf{0.960}  \\
    LSTM (semantic)            & \textbf{0.967} &\textbf{ 0.967} & 0.953$\downarrow$ \\
    LSTM (next\_log, semantic) & \textbf{0.784} & 0.781$\downarrow$ & 0.742$\downarrow$\\
    \midrule
    \multicolumn{2}{l}{\# Models degraded} & 5/7 & 5/7 \\
    \bottomrule
  \end{tabular}
\end{table}

Effective data augmentation for anomaly detection must address two key challenges: achieving sufficient coverage of normal execution paths and generating plausible parameter values. Coverage is the primary goal: augmented data must include execution paths that are absent from the original dataset, enabling models to recognize them as normal. Plausibility is the secondary requirement: although the parameter values are synthesized by an LLM rather than drawn from real logs, they must be sufficiently coherent and context-aware so that models can extract meaningful features without being misled by unrealistic artifacts.
We propose AnomalyGen, a method designed to tackle these challenges. It leverages static analysis to systematically explore the space of valid execution paths, and employs an LLM to determine which path combinations are logically feasible and to synthesize plausible parameter values. The effectiveness gains we observe come mainly from coverage: models no longer misclassify unseen yet valid execution paths as anomalies.

AnomalyGen operates in three progressive phases:
In Phase I, AnomalyGen performs log-oriented static analysis: it prunes the global call graph to retain only methods involved in log sequences, then constructs LCFGs that capture branching and sequencing constraints among log statements. This step defines the set of structurally valid execution paths that the system can produce. 
In Phase II, AnomalyGen uses a stack-based simulation to assemble cross-method log sequences, and applies an LLM with CoT reasoning to verify that each assembled path is logically consistent (a callee's preconditions are satisfied by its caller's variable state) and to generate plausible parameter values for each log template.
In Phase III, AnomalyGen automatically labels generated sequences using domain-knowledge rules based on log severity levels and semantic keywords, enabling scalable supervised training. Our key contributions are as follows:
\begin{itemize}
    \item We propose AnomalyGen, the first framework to integrate log-oriented static analysis with LLM-based CoT reasoning to synthesize parameterized log sequences for anomaly detection. The combination allows AnomalyGen to generate log sequences that follow valid control flow and contain plausible runtime parameters, which prior methods could not achieve simultaneously. To the best of our knowledge, AnomalyGen is \textbf{the first approach to automated log synthesis using LLMs for anomaly detection}.
    \item We conduct an empirical study showing that popular log benchmarks cover less than 10\% of actual system execution behaviors, and we argue that this data sparsity is the primary bottleneck for anomaly detection performance in practice. 
    \item We evaluate AnomalyGen across 12 detection models on two real-world systems, confirming AnomalyGen's effectiveness in improving model performance across different paradigms. Besides, we find consistent improvements with a maximum F1-score gain of 15.2\% on HDFS and 13.0\% on Zookeeper.  
    \item We publicly release our source code and generated datasets to facilitate reproducibility and future research.
\end{itemize}

\section{Related Work}

\subsection{Log-Based Anomaly Detection}
Log-based anomaly detection methods can be grouped into three paradigms. 
Classical machine learning approaches~\cite{10.1145/1081870.1081927, 180694} use statistical features such as event count vectors, PCA projections, or invariant relationships between event frequencies. While recent studies~\cite{yu2024deep} demonstrate their continued efficiency, these methods struggle with complex, sequential semantic patterns inherent in modern distributed systems. 
Supervised deep learning models treat anomaly detection as sequence classification: CNN-based~\cite{8511880} and LSTM-based~\cite{chen2022experiencereportdeeplearningbased} architectures learn to classify log sequences given labeled training data. Their main limitation is the requirement for labeled anomalies, which are expensive to collect in practice. 
Consequently, the field has shifted toward unsupervised deep learning methods (e.g. DeepLog~\cite{du2017deeplog} and LogAnomaly~\cite{10.5555/3367471.3367702}) that learn a model of normal behavior from unlabeled data and flag sequences that deviate from this model. More recent work has extended this paradigm with Transformers~\cite{guo2024logformer, 9244088}, graph structures~\cite{li2024graph, 10411677}, and contrastive learning~\cite{zhang2024logicodellmdrivenframeworklogical, shavit2024semantilog}. 
All three paradigms share a common dependency on training data. Supervised models need labeled anomalies; unsupervised models learn a model of normality from training logs. When training data is sparse, the learned model of normality is incomplete, and valid execution paths not seen during training appear anomalous to the model. AnomalyGen targets this shared limitation by expanding training data coverage.

\subsection{Data Augmentation for Log Analysis}
Existing augmentation approaches work either on observed log data or on source code. 
Data-based methods, such as LogRobust~\cite{10.1145/3338906.3338931}, apply transformations (insertion, deletion, shuffling) to existing log sequences. Because they cannot introduce log templates not already in the dataset, they do not address the coverage problem we identify. Methods that work on code, most prominently AutoLog~\cite{huo2023autolog}, traverse program control flow graphs to generate new sequences. AutoLog ensures that generated sequences respect the program's structural constraints, which is an important property. However, it does not verify cross-method logical consistency (e.g., whether a callee's conditions are satisfied by a given caller context), and it produces only log template sequences without parameter values. AnomalyGen extends the code-based generation approach by using LLM reasoning to handle both of these gaps.

\subsection{Log Statement Generation}
A closely related field is the automatic generation of logging statements, deciding where and what to log. Prior work relied on deep learning (LANCE~\cite{mastropaolo2022using}, LoGenText~\cite{ding2022logentext}) or static context (SCLogger~\cite{li2024go}) to recommend log placement. With the advent of LLMs, tools like UniLog~\cite{xu2024unilog} and recent empirical studies~\cite{li2024exploring} have demonstrated that LLMs possess deep semantic understanding of logging intent and code context. 
However, these works focus on generating code (injecting static logging statements), not generating log sequences (producing runtime execution traces). While they prove that LLMs understand logs, they do not address the challenge of chaining these logs into coherent, long-range sequences for anomaly detection. AnomalyGen is the first to leverage semantic understanding of LLMs to simulate the \textit{execution trajectory} of logs, rather than just their static placement.

\subsection{LLMs for Log Analysis}
Several recent systems use LLMs as inference-time components for log analysis. KnowLog~\cite{ma2024knowlog} fine-tunes language models for log parsing; LogPrompt~\cite{liu2024logprompt, liu2024interpretable} and PreLog~\cite{10.1145/3654966} use in-context learning for zero-shot anomaly detection. These systems use LLMs to analyze logs at deployment time. 
AnomalyGen uses LLMs in a different role. 
The LLM is used offline during training data construction to judge path feasibility and instantiate parameter values, while anomaly detection at deployment time is still handled by downstream models.

\section{Methodology}
\subsection{Overview}
Log anomaly detection models are fundamentally constrained by the quality of training data: real-world logs are often incomplete, environment-dependent, and difficult to label at scale. Automatically generating log sequences from source code is a natural solution, but it requires answering two practical questions: \textbf{which execution paths are valid, and what do the generated sequences look like at runtime?}
Current approaches generally fall into two categories. Methods based on fault injection or perturbation~\cite{10.1145/3338906.3338931} can capture runtime behavior but are limited to scenarios explicitly anticipated by their designers; they cannot generate execution paths that developers did not think to test. Methods based on direct CFG traversal, such as AutoLog~\cite{huo2023autolog}, can enumerate execution paths defined in the code, but they cannot resolve paths whose feasibility depends on runtime variable values, and they cannot fill in realistic log parameters. The result is sequences of log template stubs that lack the contextual information detection models need.

\begin{figure*}
  \includegraphics[width=\textwidth]{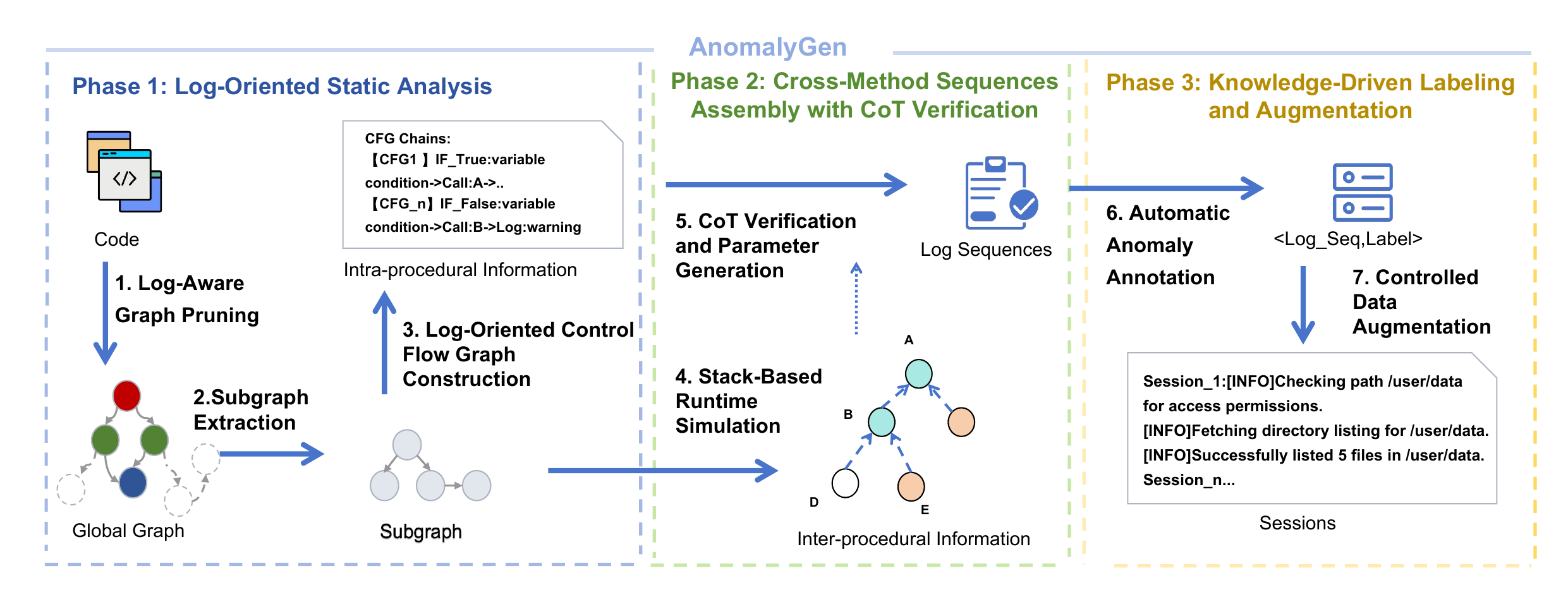}
  \caption{The three-stage workflow of AnomalyGen. Phase I performs log-based graph pruning (1), subgraph extraction (2), and LCFG construction (3) to capture intra-procedural logic. Phase II reconstructs inter-procedural paths via stack-based simulation (4) and merges log sequences with CoT semantic verification (5). Phase III completes the pipeline with automatic anomaly annotation (6) and controlled data augmentation (7) to produce high-quality labeled sessions.}
  \label{fig:Framework}
\end{figure*}

AnomalyGen addresses these limitations through a three-phase pipeline (Figure~\ref{fig:Framework}). Phase I uses static analysis to prune the call graph and construct LCFGs, defining a space of structurally valid log sequences. Phase II uses stack-based simulation and LLM-based reasoning to assemble cross-method sequences and verify their logical consistency, while also generating plausible parameter values. Phase III applies knowledge-driven rules to label each generated sequence as normal or anomalous, producing training data ready for downstream models.

\subsection{Phase I: Log-Oriented Static Analysis}
The goal of Phase I is to identify which methods in a codebase participate in log emission, and to build a control flow representation that captures how logging statements are ordered and constrained within each method. We do this in three steps.
\subsubsection{Global Call Graph Pruning}
A software system can be modeled as a directed call graph $G = (V, E)$ where $V$ is the set of methods and $E$ records invocation relationships. In large systems, only a small fraction of methods directly emit logs ($V_{log} \subset V_{total}$). Traversing the full call graph is both wasteful and counterproductive, because most paths lead to no log output and generating an exponentially growing set of paths~\cite{nejmeh1988npath} is irrelevant to our goal.

\begin{figure}[!htb]
    \centering
    \includegraphics[width=0.4\textwidth]{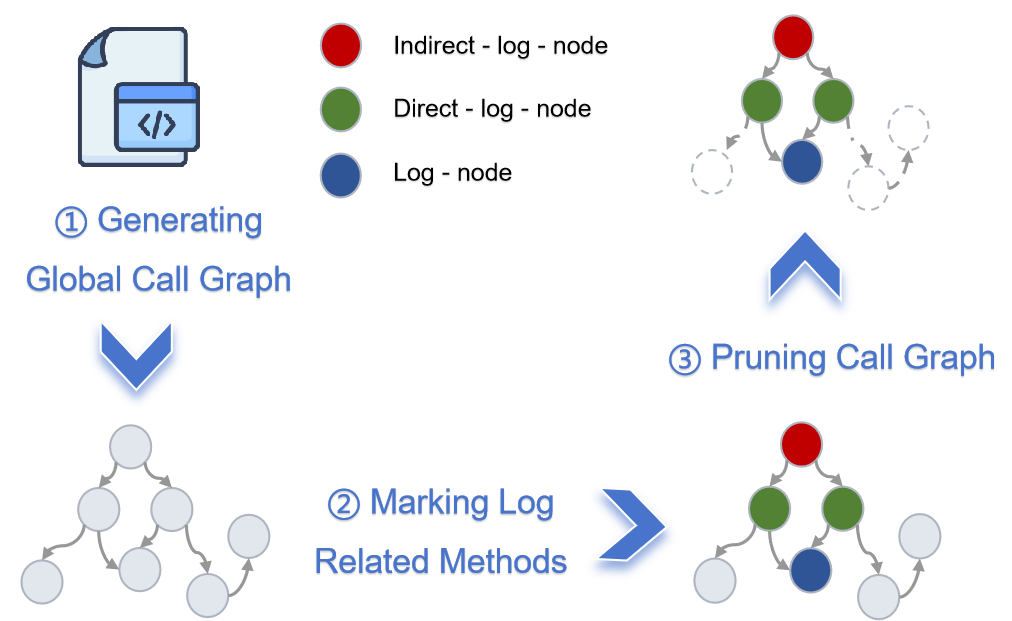}
    \caption{Log-aware node labeling and call graph pruning.}
    \label{fig:log_pruning} 
\end{figure}

To address this, we reformulate the problem not as reachability from a starting point, but as \textit{backward reachability to a log statement}: a node $v$ is relevant if and only if there exists a path from $v$ to a logging API call. We define two types of log-relevant methods: \textit{anchor nodes (Direct\_Log\_Node)}, which directly call logging APIs (e.g., \texttt{org.slf4j.Logger.info}), and \textit{transitive nodes (Indirect\_Log\_Node)}, which are callers that eventually reach an anchor node. We compute transitive nodes by inverting the edge direction to form $E^{-1}$ and traversing from the set of anchor nodes. Any method not reached in this reverse traversal is guaranteed to produce no log output and is pruned. As illustrated by the pruning process in Figure~\ref{fig:log_pruning}, our analysis of Hadoop v3.3.6 reduced the call graph to 15.11\% of its original size.

\subsubsection{Subgraph Extraction}

After pruning, the remaining graph is still a large connected structure containing call chains of arbitrary depth. Feeding such chains into Phase II would cause two problems. First, deep traversal of a large graph still produces an unmanageable number of paths. Second, very long call chains produce code contexts that exceed the effective window of the LLM used for reasoning in Phase II, degrading the quality of its verification.

We address this by partitioning the pruned graph into manageable subgraphs using two thresholds. The entry threshold $T_{entry}$ controls how many entry points are selected for analysis, keeping focus on the most representative invocation paths. The depth threshold $T_{depth}$ limits the traversal depth of each call chain, ensuring that the code context for any subgraph fits within the LLM's effective context (e.g., matching the token limit of models like GPT-4 or Llama-3). Each resulting subgraph contains at least one log statement and represents a bounded execution unit. This partitioning balances the coverage of execution paths against the computational cost of Phase II analysis.

\begin{figure*}[!htb]
    \centering
    \includegraphics[width=1\textwidth]{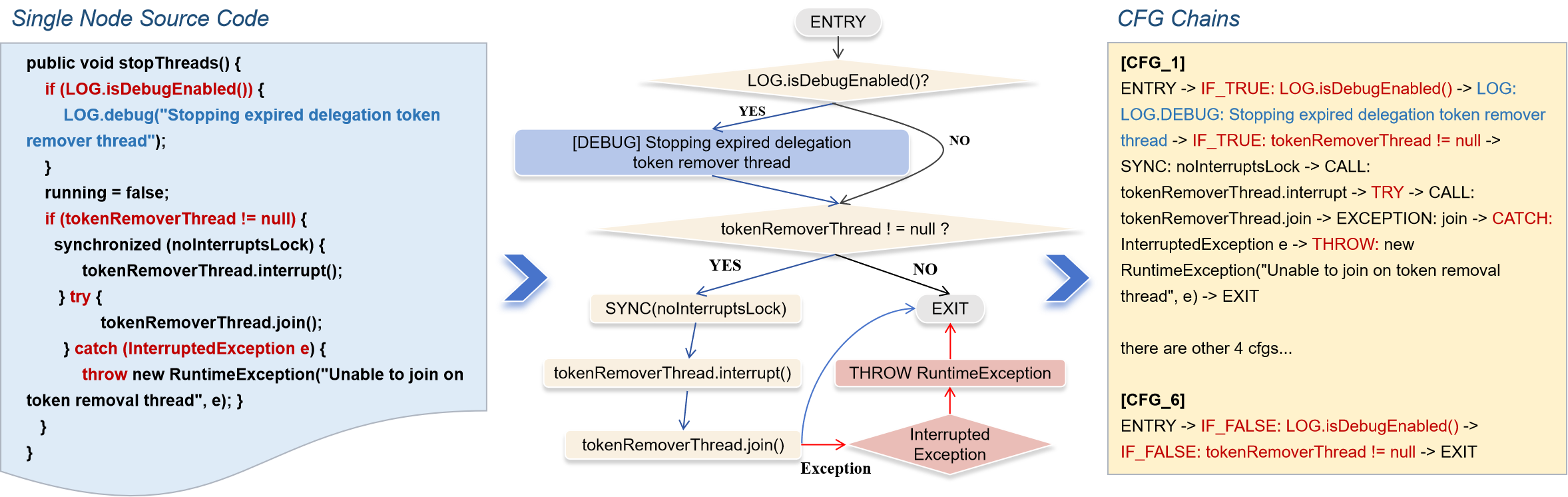}
    \caption{An example of LCFG construction: AST analysis extracts log-critical elements, and dominance analysis determines the valid ordering constraints.}
    \label{fig:lcfg}
\end{figure*}

\subsubsection{Log-Oriented Control Flow Graph Construction}
The pruned call graph identifies methods involved in logging, but it does not describe the ordering of log statements within a method. 
A standard control flow graph contains instruction-level detail (arithmetic operations, variable assignments) that is irrelevant to log sequencing. As shown in Figure~\ref{fig:lcfg}, we construct a Log-Oriented Control Flow Graph (LCFG) that retains only the information needed to determine valid log orderings.

To bridge this gap, we construct a specialized LCFG, a reduced control flow representation that explicitly maps program logic to log outputs. LCFG construction proceeds in three steps. We first use AST analysis to extract the log-critical elements of each method body: log emission points, control flow structures (conditionals, loops, switch statements), and outgoing method calls. We then apply dominance analysis to determine ordering relationships among log statements. If log statement $A$ dominates log statement $B$ (i.e., every execution path to $B$ passes through $A$), they are placed in strict sequential order $A \to B$. If two log statements appear in mutually exclusive branches (e.g., the then-branch and else-branch of a conditional), they are placed on separate disjoint paths in the LCFG, preventing AnomalyGen from generating sequences where both appear together. Finally, we annotate LCFG edges at method call sites with call-site constraints, linking the LCFG to the call graph for use in Phase II's cross-method assembly.

\begin{figure*}[!htb]
    \centering
    \includegraphics[width=0.9\textwidth]{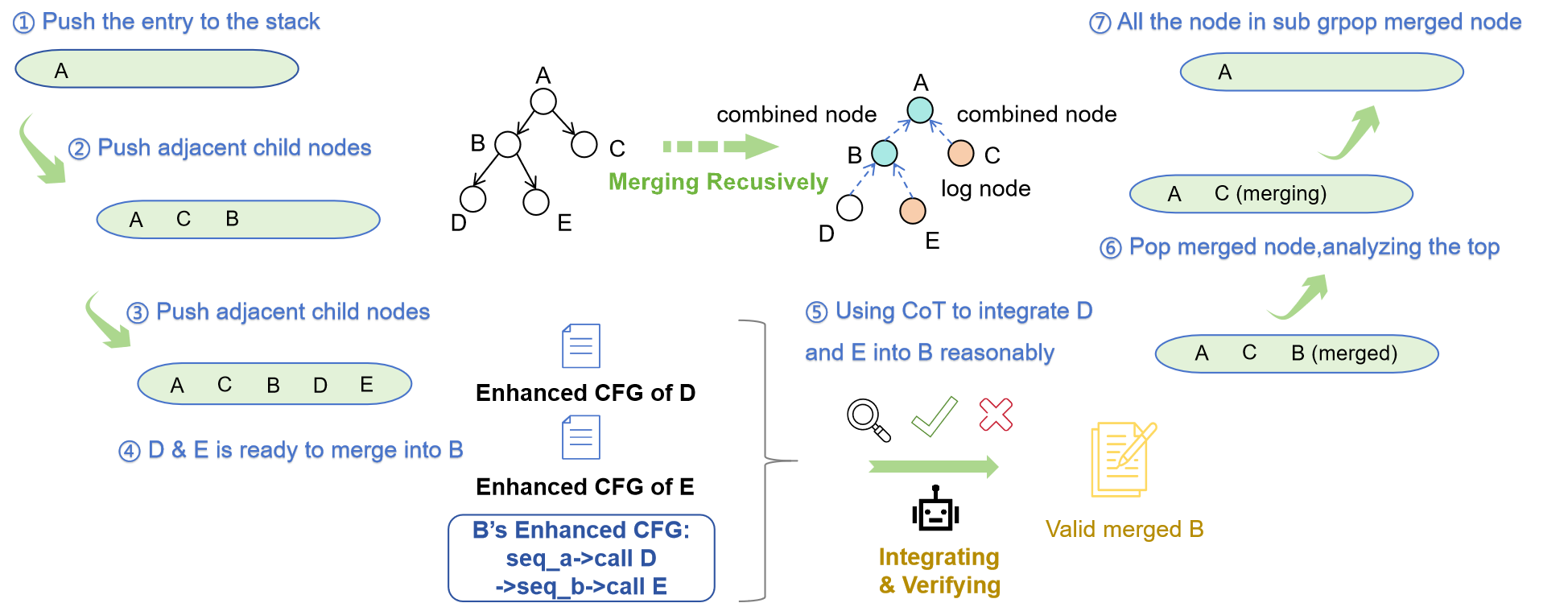}
    \caption{Phase II: Recursive Log Merging with CoT Verification Process.}
    \label{fig:recursive_merge}
\end{figure*}

\subsection{Phase II: Cross-Method Sequence Assembly with CoT Verification.}
Phase I produces a set of LCFGs that capture intra-method log ordering constraints. Phase II assembles these into complete, cross-method log sequences. 
The main challenge is that a callee path may be structurally reachable yet infeasible under the caller state. 
Whether a branch is taken may depend on argument values, flags, or prior control flow.
For example, whether a flag was set, whether a pointer is null, or whether a loop has been entered. Static structure alone cannot determine whether a given combination of caller and callee execution paths is logically possible. To bridge the gap between static structure and dynamic reality, we implement a virtual stack that mimics the relationship across methods, use CoT to verify logic consistency and generate parameters.

\subsubsection{Stack-Based Runtime Simulation}
Real program execution follows a call stack: when a method is invoked, execution enters the callee; when the callee returns, execution resumes in the caller. AnomalyGen mimics this using a virtual stack. When traversal of a caller's LCFG reaches a method call site, the current execution context (variable states and path constraints from the caller) is pushed onto the stack. Traversal then descends into the callee's LCFG to generate its log sub-sequence. On return, the stack is popped and the callee's sub-sequence is appended to the caller's ongoing sequence. This process, as shown in Figure~\ref{fig:recursive_merge}, is applied recursively, following the call graph structure identified in Phase I.

The result of this step is a set of candidate sequences that are structurally assembled from the LCFGs of multiple methods. However, structural consistency does not guarantee logical consistency: a callee may require preconditions (e.g., \texttt{arg != null}) that the caller does not satisfy for a given path. Simply concatenating paths without checking this would generate sequences that are structurally valid but represent impossible executions, which would corrupt training data. We use an LLM to resolve this.


\begin{figure}[h]
\centering
\includegraphics[width=1.0\linewidth]{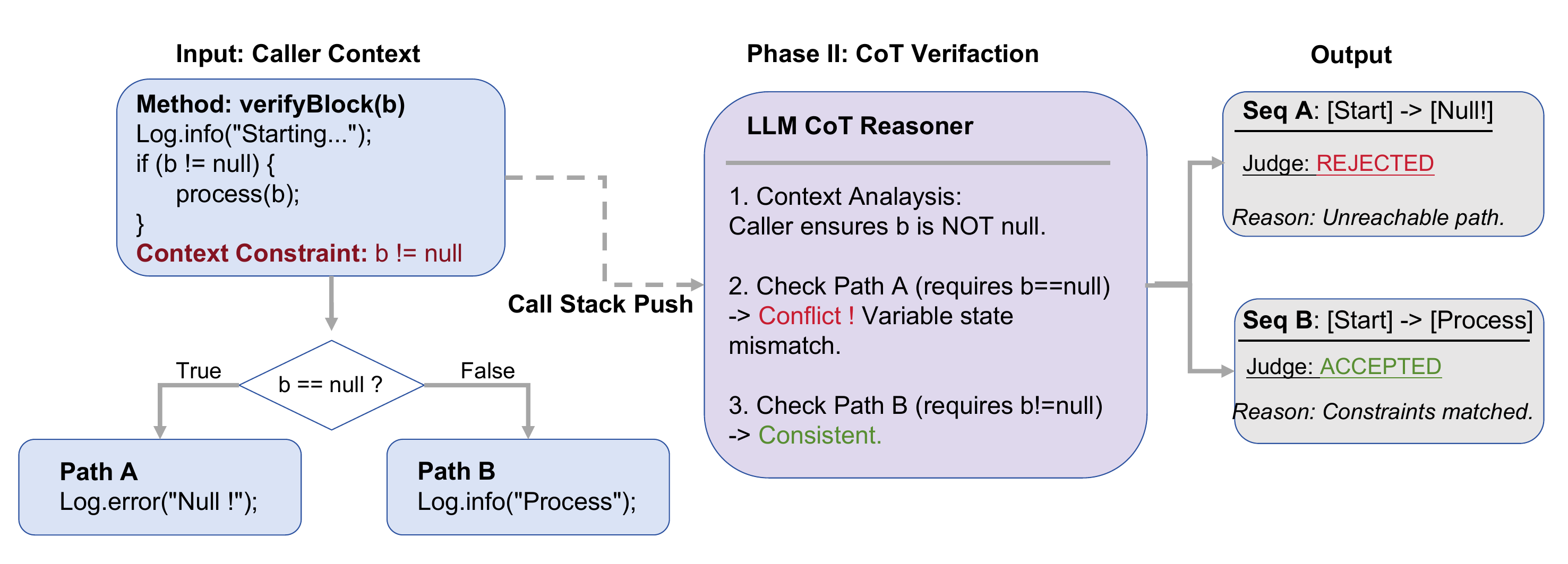}
\caption{Example of Phase II: a caller context and candidate callee path are provided to the LLM, which evaluates logical consistency and generates parameter values.}
\label{fig:cot_verification}
\end{figure}

\subsubsection{CoT Verification and Parameter Generation}
\begin{figure}[h]
\centering
\includegraphics[width=0.8\linewidth]{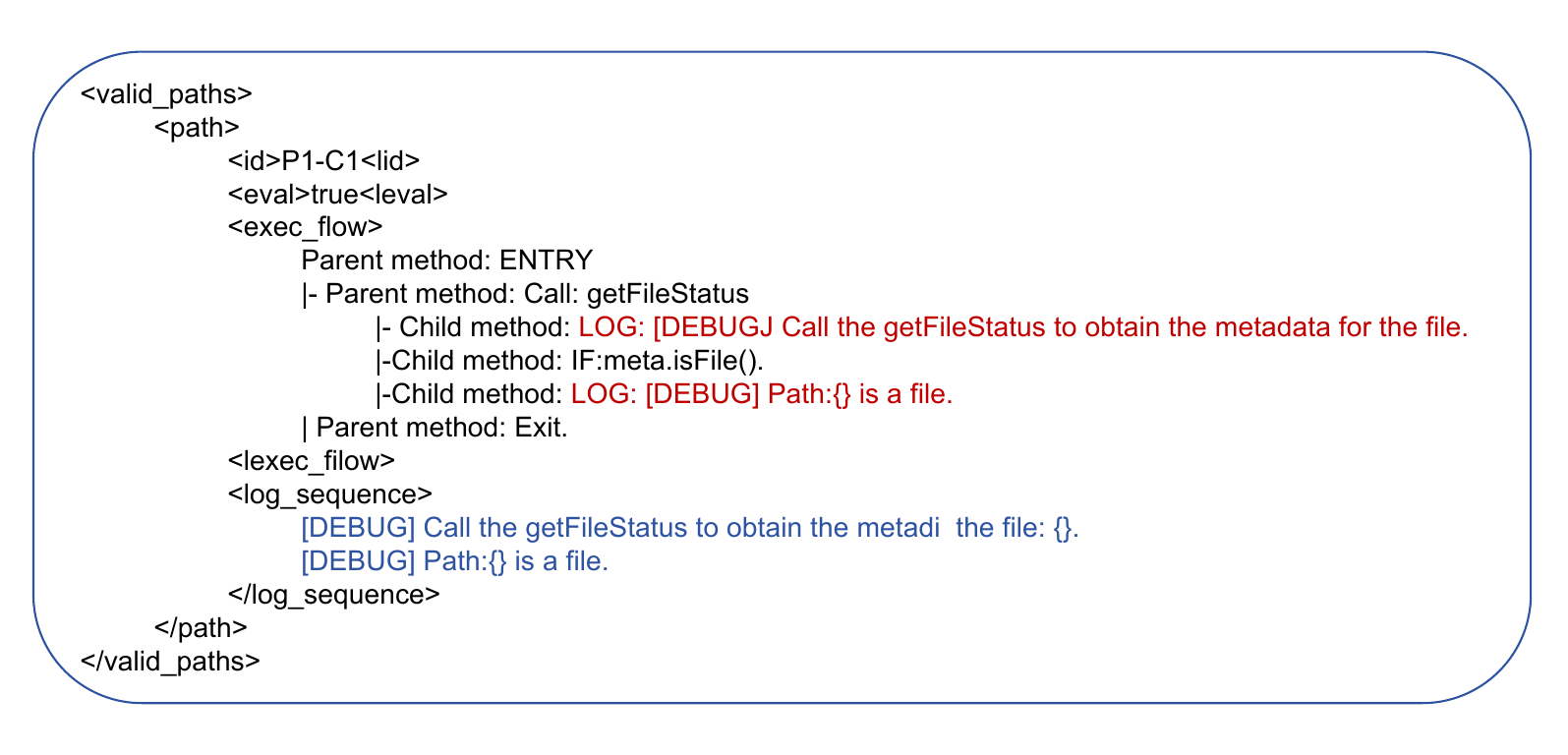}
\caption{Structured output produced by CoT verification.}
\label{fig:structured output}
\end{figure}

To check logical consistency and generate parameters, we use an LLM with Chain-of-Thought prompting (Figure~\ref{fig:cot_verification}). At each merge point where a callee sub-sequence is about to be appended to a caller path, we construct a structured prompt containing: (1) the source code of the calling method, highlighting argument definitions and branch conditions; (2) the candidate callee path segment; and (3) static analysis hints, such as type constraints on parameters. The LLM reasons step by step about whether the callee path is consistent with the caller context. For example, checking whether a null-check in the callee would be satisfied given the arguments the caller is passing, or whether a condition in the callee is reachable given the caller's variable state. The output is a structured JSON object (Figure~\ref{fig:structured output}) containing a boolean validity judgment and a brief rationale. Candidate paths judged invalid are discarded.

For valid paths, we prompt the LLM to generate plausible values for the variable portions of each log template. Rather than leaving templates as stubs (e.g., \texttt{blk\_<*>}), the LLM fills in contextually appropriate values (e.g., a realistic block ID) by reasoning about the code context. This step is important for semantic-encoding detection models, which rely on the content of log messages rather than just their identity.

This phase successfully integrates log information from source code through recursive merging and CoT verification, reconstructing log sequences that match actual execution flows and providing high-quality training data for downstream anomaly detection models.

\subsection{Phase III: Knowledge-Driven Labeling and Augmentation}
Phase II produces a set of log sequences that follow valid control flow and contain plausible parameters. The remaining task is to assign anomaly labels for supervised training and to integrate the synthetic sequences with real training data in a controlled way. Manual annotation is unscalable, and random labeling introduces noise. AnomalyGen adopts a knowledge-driven strategy, grounded in industrial software reliability engineering practices.

\subsubsection{Automatic Anomaly Annotation}

We use two types of evidence to assign labels. First, we check logging severity levels: sequences containing \texttt{ERROR} or \texttt{FATAL} level messages, or sequences that include explicit exception class names (e.g., \texttt{IOError}, \texttt{NullPointerException}), are labeled anomalous. This follows standard practice in reliability engineering, where developers rely on severity levels as a primary signal when inspecting logs for failures~\cite{7774521}. Second, for failures that do not produce high-severity messages (e.g., a timeout handled internally), we apply configurable keyword matching on message content using terms such as ``timeout'', ``refused'', and ``invalid state'', as well as HTTP/RPC status code patterns. Sequences that meet neither criterion are labeled normal.

To assess labeling accuracy, we manually reviewed a random sample of generated sessions. Two domain experts independently checked each session's label against the execution context. For Zookeeper, we reviewed 141 sessions (360 log lines) and found 6 mislabeled sessions, giving a labeling accuracy of \textbf{95.74\%}. For HDFS, we reviewed 106 sessions (287 log lines) and found 5 mislabeled sessions, giving an accuracy of \textbf{95.28\%}. Analysis of the mislabeled cases reveals three recurring causes. Some anomaly indicators use vague identifiers not captured by our keyword set (e.g., \texttt{[WARN] Unknown packet type}). Some failures are implied by state transitions rather than explicit error messages (e.g., a debug log \texttt{Socket is not open} followed by \texttt{Closing socket connection} indicates a failure but contains no error-level log). And some recovery actions (e.g., \texttt{[INFO] Restarting node due to timeout}) were labeled normal because the triggering failure was not visible in the generated sequence.

One advantage of this labeling strategy is that it interacts with Phase II's path generation. Because AnomalyGen generates contextually coherent sequences, an anomaly label does not correspond to an isolated error message but to a complete execution trace leading to a failure. This gives detection models training examples that capture the sequential context of anomalies, rather than just their surface markers. Finally, we consolidate the generated data into triplets of \texttt{<Log Sequence, Execution Context, Anomaly Label>}.

\subsubsection{Controlled Data Augmentation Strategy}
To preserve evaluation integrity, we strictly adhere to two protocols of integrating synthetic data into training. To prevent data leakage, we strictly segregate the dataset. Synthetic data is exclusively injected into the training set; the test set remains composed entirely of original, real-world log sequences. To control augmentation volume, we define a ``session'' as the fundamental unit of augmentation (a complete log sequence associated with a task ID). Then we define an augmentation ratio $R = N_{syn} / N_{real}$ and sample from the synthetic pool to achieve the desired ratio. We balance normal and anomalous sequences in the augmented training set to mitigate class imbalance.

\section{Evaluation}
To validate the effectiveness, robustness, and practicality of AnomalyGen, we conduct comprehensive experiments to answer the following research questions (RQs):
\begin{itemize}
    \item RQ1: How effective is AnomalyGen in improving model performance across different paradigms?
    \item RQ2: What is the contribution of each core component of AnomalyGen?
    \item RQ3: How do different experimental settings impact AnomalyGen's effectiveness?
    \item RQ4: What is the computational overhead of AnomalyGen?
\end{itemize}

\subsection{Experiment Setup}
\subsubsection {Implementation}
AnomalyGen is implemented in Java and Python. We use java-callgraph2~\cite{java-callgraph2} for call graph generation and JavaParser~\cite{javaparser_2020_3842713} for AST analysis. For LLM-based reasoning, we employ GPT-4o~\cite{openai2024gpt4ocard} and DeepSeek-V3~\cite{deepseekai2024deepseekv3technicalreport}, both of which demonstrate strong code understanding and reliable CoT-based reasoning~\cite{wei2023chainofthoughtpromptingelicitsreasoning}. We set the temperature to 0 for all LLM calls to ensure that our augmented datasets can be exactly reproduced. All anomaly detection models are evaluated on identical augmented data, making performance comparisons directly attributable to model characteristics rather than data variation. To ensure reproducibility and enable fair comparison across different detection models, we adopt two key practices: Firstly, we set the temperature parameter to 0 for all LLM API calls during data generation, ensuring that identical inputs always produce identical outputs. Secondly, we evaluate all anomaly detection models on identical augmented data, making performance comparisons directly attributable to model characteristics rather than data variation.

\subsubsection {Experimental Environment}
Our experiments were conducted on a server equipped with a dual Intel (R) Xeon (R) Platinum 8336C CPU@2.30GHz (32 cores total) and 192GB of RAM, and NVIDIA GeForce RTX 4090 (2 GPUs, 24 GB).

\subsubsection {Anomaly Detection Models}
To ensure a comprehensive evaluation, we select 12 representative anomaly detection models spanning four distinct paradigms, following recent empirical studies \cite{yu2024deep, chen2022experiencereportdeeplearningbased}. This allows us to assess the impact of AnomalyGen across a wide range of architectural assumptions and learning strategies:

\begin{itemize}
    \item \textbf{Classic Machine Learning Models.} This category represents the foundational approaches in log anomaly detection, which rely on simple, interpretable feature-based learning and serve as important baselines for evaluating more complex methods. We evaluate three representative models: Decision Tree (DT)~\cite{1301345}, Single-Layer Feedforward Neural Network (SLFN)~\cite{10.1109/72.846750}, and k-Nearest Neighbors (KNN)~\cite{Fix1989DiscriminatoryA}.
    \item \textbf{Supervised Deep Learning Models.} This paradigm represents scenarios where labeled training data is available, enabling models to learn explicit mappings from log sequences to anomaly labels. We select models that treat anomaly detection as a sequence classification task, requiring labeled data. We evaluate a Convolutional Neural Network (CNN) based on the architecture proposed by~\cite{8511880} and a supervised Long Short-Term Memory (LSTM) network inspired by~\cite{10.1145/3338906.3338931}.
    \item \textbf{Unsupervised Deep Learning Models.} This category represents the most common scenario in practice, where models must learn from unlabeled data. We evaluate two prominent models under the next-log prediction paradigm: an LSTM based on the seminal DeepLog~\cite{du2017deeplog} and a Transformer model adapted for log anomaly detection~\cite{9338283}. Additionally, we evaluate an unsupervised LSTM leveraging sequential and quantitative anomalies~\cite{10.5555/3367471.3367702} in unstructured logs.
    \item \textbf{LLM-based Model.} To assess our framework against the latest paradigm, we evaluate PreLog~\cite{10.1145/3654966}, which utilizes in-context learning capabilities of LLMs to detect anomalies by analyzing the semantic plausibility of log sequences.
\end{itemize}
All deep learning models are implemented with the Deep-Loglizer toolkit~\cite{chen2022experiencereportdeeplearningbased} and classical ML models with LightAD~\cite{yu2024deep}. For each deep learning model, we evaluate two input encoding schemes: a sequential encoding based on log event IDs, and a semantic encoding based on log message text. Comparing these two different encoding schemes allows us to analyze the interplay between our data augmentation approach and different feature extraction approaches. To maintain the comparability of results, we guarantee consistency across test sets under different augmentation ratios and models. For deep learning models, we retain the original 8:2 split ratio. For classical machine learning, we employ a 7:3 dataset split to better observe the dataset's effectiveness~\cite{chen2022experiencereportdeeplearningbased}. 

\subsubsection {Metrics}
To evaluate the accuracy and effectiveness of anomaly detection techniques, we employ Precision, Recall and F1-score as evaluation metrics. These metrics are calculated based on the number of True Positives (TP), False Positives (FP), and False Negatives (FN), where \textit{positive} refers to an anomaly.
\begin{itemize}
\item \textbf{Precision.}
It measures the accuracy of the positive predictions. High precision indicates a low false alarm rate.
\begin{equation}
\text{Precision} = \frac{\text{TP}}{\text{TP} + \text{FP}}
\end{equation}
\item \textbf {Recall.} 
It measures the model's ability to identify all actual positive instances. High recall indicates a low rate of missed anomalies.
\begin{equation}
    \text{Recall} = \frac{\text{TP}}{\text{TP} + \text{FN}}
\end{equation}
\item \textbf{F1-score.}
It is the harmonic mean of Precision and Recall. It provides a single score that balances both metrics, where simply maximizing either precision or recall can be misleading. A high F1-score (close to 1) indicates that the model achieves both high precision and high recall, meaning it accurately identifies most actual anomalies while minimizing false alarms—this is the ideal performance for anomaly detection tasks. In contrast, a low F1-score (close to 0) signifies that the model fails to balance precision and recall: it may either miss a large number of real anomalies (low recall), generate excessive false alarms (low precision), or suffer from both issues simultaneously, leading to poor overall anomaly detection performance.
\begin{equation}
    \text{F1-score} = 2 \times \frac{\text{Precision} \times \text{Recall}}{\text{Precision} + \text{Recall}}
\end{equation}
\end{itemize}
In the following experiments, we use the abbreviation F1 to denote F1-score, RC to denote Recall, and PC to denote Precision.

\subsection {Dataset}
To ensure a fair and reproducible comparison with prior work~\cite{du2017deeplog, li2020swisslog,10.1145/3338906.3338931,yang2021plelog,le2022log,yu2024deep}, we use public dataset released from the widely-adopted LogHub benchmark suite~\cite{zhu2023loghub}. 
These are the two labeled datasets in LogHub whose Java source code is publicly accessible, which is required for AnomalyGen's static analysis. 
HDFS is a widely-used benchmark for anomaly detection research~\cite{du2017deeplog, li2020swisslog,10.1145/3338906.3338931,yang2021plelog,le2022log,yu2024deep}; it consists of 11,175,629 log messages collected on Amazon EC2, segmented into sequences by block ID.
Another system, Zookeeper, provides critical distributed synchronization and coordination services, making it another representative and practically relevant benchmark for anomaly detection research~\cite{huo2023autolog,10.1145/3441448,duan2025logactionconsistentcrosssystemanomaly}. The dataset was collected from a 32-node lab cluster over 26.7 days and contains approximately 74,380 log messages covering behaviors such as leader election and session management~\cite{huo2023autolog,10.1145/3441448,duan2025logactionconsistentcrosssystemanomaly}. 

To compare the diversity of baseline and augmentation dataset, we take the number of log events and coverage of log records into account. As shown in Table~\ref{tab:log_coverage}, \textit{LHub-HDFS} presents baseline, while \textit{AG-HDFS} is the dataset generated by AnomalyGen. Compared to existing logging datasets, AnomalyGen effectively enhances the comprehensiveness of the HDFS' logging events, achieving coverage of about 99.48\%. The number of logging events generated by AnomalyGen is 95 times higher than that of traditional datasets (the HDFS scenario scales 95 times from 30 to 2874 events). In addition, in manual validation with randomly sampled data, we conduct a detailed validation analysis, and the experimental results show that AnomalyGen covers most of the datasets in the existing data, with a ratio of 14/15 for HDFS, which indirectly guarantees the authenticity of the log.

\begin{table*}[!tb]
    \caption{A comparison of datasets for baseline dataset and augmentation dataset on HDFS.}
    \label{tab:log_coverage}
    \centering
        \begin{tabular}{c|ccccccc|ccccccc}
        \hline
            &Dataset & \# Log Events & Logging Coverage & R-Coverage & Improvement\\ 
        \hline
        HDFS & LHub-HDFS & 30 & 1.04\% (30/2889)   & 93.33\% (14/15) & 191 $\times$ \\
        - & AG-HDFS & 2874 & 99.48\% (2874/2889)   & - & - \\
        \hline
        \end{tabular}
\end{table*}

\begin{table*}[tbp]
  \centering
  \caption{Overall evaluation performance on HDFS. Bold indicates the best performance, and underlined signifies better than the baseline.}
  \label{tab:evaluation_performance}
  \resizebox{\textwidth}{!}{%
  \begin{tabular}{@{}ll*{15}{c}@{}}
    \toprule
    \multirow{3}{*}{Model} & \multirow{3}{*}{Method} & \multicolumn{15}{c}{Anomaly Ratio} \\
    \cmidrule(lr){3-17}
    & & \multicolumn{3}{c}{0.0} & \multicolumn{3}{c}{0.001} & \multicolumn{3}{c}{0.01} & \multicolumn{3}{c}{0.1} & \multicolumn{3}{c}{1.0} \\
    \cmidrule(lr){3-5} \cmidrule(lr){6-8} \cmidrule(lr){9-11} \cmidrule(lr){12-14} \cmidrule(lr){15-17}
    {}& {} & F1 & RC & PC & F1 & RC & PC & F1 & RC & PC & F1 & RC & PC & F1 & RC & PC \\
    \midrule
    
    \multirow{2}{*}{CNN} & sequentials & 0.961 & 0.995 & 0.929 & \underline{\textbf{0.972}} & \underline{\textbf{0.995}} & \underline{\textbf{0.949}} & \underline{0.970} & \underline{0.995} & \underline{0.946} & \underline{0.962} & \underline{0.995} & \underline{0.931} & \underline{0.971} & \underline{0.995} & \underline{0.948} \\
    & semantics & 0.964 & 0.988 & 0.942 & \underline{\textbf{0.971}} & \underline{\textbf{0.989}} & \underline{\textbf{0.954}} & \underline{0.969} & \underline{0.984} & \underline{0.953} & \underline{0.967} & \underline{0.988} & \underline{0.948} & \underline{0.967} & \underline{0.988} & \underline{0.947} \\
    \midrule
    \multirow{2}{*}{Transformer} & next\_log & 0.818 & 0.829 & 0.807 & \underline{0.956} & \underline{0.963} & \underline{0.949} & \underline{0.899} & \underline{0.928} & \underline{0.872} & \underline{\textbf{0.970}} & \underline{\textbf{0.925}} & \underline{\textbf{0.820}} & \underline{0.946} & \underline{0.964} & \underline{0.929} \\
    & semantics & 0.727 & 0.605 & 0.911 & \underline{0.691} & \underline{0.573} & \underline{0.872} & \underline{0.721} & \underline{0.580} & \underline{0.951} & \underline{\textbf{0.737}} & \underline{\textbf{0.591}} & \underline{\textbf{0.979}} & \underline{0.701} & \underline{0.573} & \underline{0.903} \\
    \midrule
    \multirow{3}{*}{LSTM} & sequentials (sup) & 0.959 & 0.995 & 0.925 & \underline{\textbf{0.960}} & \underline{\textbf{0.990}} & \underline{\textbf{0.931}} & \underline{0.950} & \underline{0.996} & \underline{0.908} & \underline{\textbf{0.960}} & \underline{\textbf{0.996}} & \underline{\textbf{0.928}} & \underline{\textbf{0.960}} & \underline{\textbf{0.995}} & \underline{\textbf{0.926}} \\
    & next\_log (unseq) & 0.919 & 0.954 & 0.886 & \underline{0.938} & \underline{0.910} & \underline{0.967} & \underline{0.935} & \underline{0.910} & \underline{0.961} & \underline{0.928} & \underline{0.899} & \underline{0.960} & \underline{\textbf{0.949}} & \underline{\textbf{0.939}} & \underline{\textbf{0.959}} \\
    & semantics & 0.967 & 0.995 & 0.941 & \underline{0.960} & \underline{0.973} & \underline{0.947} & \underline{\textbf{0.974}} & \underline{\textbf{0.989}} & \underline{\textbf{0.958}} & \underline{0.967} & \underline{0.983} & \underline{0.951} & \underline{0.968} & \underline{0.988} & \underline{0.948} \\
    \midrule
    \multirow{3}{*}{Single Method} & DT & 0.99815 & 0.99855 & 0.99764 & \underline{0.99859} & \underline{0.99879} & \underline{0.99839} & \underline{0.99859} & \underline{0.99879} & \underline{0.99839} & \underline{\textbf{0.99869}} & \underline{\textbf{0.99899}} & \underline{\textbf{0.99839}} & \underline{0.99859} & \underline{0.99879} & \underline{0.99839} \\
    & SLFN & 0.996 & 0.996 & 0.995 & \underline{0.996} & \underline{0.996} & \underline{0.996} & \underline{\textbf{0.997}} & \underline{\textbf{0.997}} & \underline{\textbf{0.997}} & \underline{0.996} & \underline{0.994} & \underline{0.997} & \underline{0.996} & \underline{0.997} & \underline{0.996} \\
    & KNN & 0.99501 & 0.99320 & 0.99683 & \underline{\textbf{0.99536}} & \underline{\textbf{0.99356}} & \underline{\textbf{0.99717}} & \underline{\textbf{0.99536}} & \underline{\textbf{0.99356}} & \underline{\textbf{0.99717}} & \underline{\textbf{0.99536}} & \underline{\textbf{0.99356}} & \underline{\textbf{0.99717}} & \underline{\textbf{0.99536}} & \underline{\textbf{0.99356}} & \underline{\textbf{0.99717}} \\
     \midrule
    \multirow{1}{*}{LLM-based} & PreLog & 0.989 & 0.989 & 0.989 & \underline{0.990} & \underline{0.991} & \underline{0.990} & \underline{\textbf{0.992}} & \underline{\textbf{0.992}} & \underline{\textbf{0.992}} & 0.977 & 0.977 & 0.969 & 0.984 & 0.984 & 0.981 \\
    \bottomrule
    
  \end{tabular}
  }
\end{table*}

\begin{table*}[tbp]
  \centering
  \caption{Overall Evaluation Performance on Zookeeper. Bold indicates the best performance, and underlined signifies better than the baseline.}
  \label{tab:evaluation_performance_zookeeper}
  \resizebox{\textwidth}{!}{%
  \begin{tabular}{@{}ll*{15}{c}@{}}
    \toprule
    \multirow{3}{*}{Model} & \multirow{3}{*}{Method} & \multicolumn{15}{c}{Anomaly Ratio} \\
    \cmidrule(lr){3-17}
    & & \multicolumn{3}{c}{0.0} & \multicolumn{3}{c}{0.001} & \multicolumn{3}{c}{0.01} & \multicolumn{3}{c}{0.1} & \multicolumn{3}{c}{1.0} \\
    \cmidrule(lr){3-5} \cmidrule(lr){6-8} \cmidrule(lr){9-11} \cmidrule(lr){12-14} \cmidrule(lr){15-17}
    {}& {} & F1 & RC & PC & F1 & RC & PC & F1 & RC & PC & F1 & RC & PC & F1 & RC & PC \\
    \midrule

    \multirow{2}{*}{CNN}
      & sequentials
      & 0.971 & 0.971 & 0.971
      & 0.967 & 0.948 & 0.988
      & 0.968 & 0.965 & 0.971
      & \underline{\textbf{0.979}} & \underline{\textbf{0.965}} & \underline{\textbf{0.994}}
      & \underline{0.974} & \underline{0.965} & \underline{0.982} \\
      & semantics
      & 0.979 & 0.959 & 1.000
      & \underline{\textbf{0.982}} & \underline{\textbf{0.965}} & \underline{\textbf{1.000}}
      & \underline{\textbf{0.982}} & \underline{\textbf{0.965}} & \underline{\textbf{1.000}}
      & 0.979 & 0.965 & 0.994
      & 0.979 & 0.965 & 0.994 \\
    \midrule

    \multirow{2}{*}{Transformer}
      & next\_log
      & 0.610 & 0.727 & 0.525
      & \underline{\textbf{0.660}} & \underline{\textbf{0.826}} & \underline{\textbf{0.550}}
      & \underline{0.612} & \underline{0.715} & \underline{0.535}
      & 0.607 & 0.791 & 0.493
      & 0.589 & 0.750 & 0.485 \\
      & semantics
      & 0.644 & 0.767 & 0.555
      & \underline{\textbf{0.660}} & \underline{\textbf{0.826}} & \underline{\textbf{0.550}}
      & \underline{0.652} & \underline{0.791} & \underline{0.555}
      & 0.613 & 0.727 & 0.530
      & 0.617 & 0.744 & 0.527 \\
    \midrule

    \multirow{4}{*}{LSTM}
      & next
      & 0.624 & 0.709 & 0.557
      & \underline{0.626} & \underline{0.686} & \underline{0.576}
      & \underline{0.639} & \underline{0.715} & \underline{0.577}
      & 0.621 & 0.703 & 0.555
      & \underline{\textbf{0.705}} & \underline{\textbf{0.791}} & \underline{\textbf{0.636}} \\
      & sequentials
      & 0.979 & 0.965 & 0.994
      & \underline{\textbf{0.982}} & \underline{\textbf{0.965}} & \underline{\textbf{1.000}}
      & \underline{\textbf{0.982}} & \underline{\textbf{0.965}} & \underline{\textbf{1.000}}
      & \underline{\textbf{0.982}} & \underline{\textbf{0.965}} & \underline{\textbf{1.000}}
      & 0.979 & 0.965 & 0.994 \\
      & next\_log (semantic)
      & \textbf{0.982} & \textbf{0.965} & \textbf{1.000}
      & \textbf{0.982} & \textbf{0.971} & \textbf{0.994}
      & \textbf{0.982} & \textbf{0.965} & \textbf{1.000}
      & \textbf{0.982} & \textbf{0.965} & \textbf{1.000}
      & 0.971 & 0.965 & 0.976 \\
      & semantics
      & 0.685 & 0.738 & 0.638
      & \underline{\textbf{0.727}} & \underline{\textbf{0.797}} & \underline{\textbf{0.668}}
      & 0.640 & 0.733 & 0.568
      & \underline{0.713} & \underline{0.808} & \underline{0.638}
      & 0.588 & 0.808 & 0.462 \\
    \midrule

    \multirow{3}{*}{Single Method}
      & DT
      & 1.000 & 1.000 & 1.000
      & \textbf{1.000} & \textbf{1.000} & \textbf{1.000}
      & \textbf{1.000} & \textbf{1.000} & \textbf{1.000}
      & \textbf{1.000} & \textbf{1.000} & \textbf{1.000}
      & \textbf{1.000} & \textbf{1.000} & \textbf{1.000} \\
      & SLFN
      & 1.000 & 1.000 & 1.000
      & \textbf{1.000} & \textbf{1.000} & \textbf{1.000}
      & \textbf{1.000} & \textbf{1.000} & \textbf{1.000}
      & \textbf{1.000} & \textbf{1.000} & \textbf{1.000}
      & \textbf{1.000} & \textbf{1.000} & \textbf{1.000} \\
      & KNN
      & 1.000 & 1.000 & 1.000
      & \textbf{1.000} & \textbf{1.000} & \textbf{1.000}
      & \textbf{1.000} & \textbf{1.000} & \textbf{1.000}
      & \textbf{1.000} & \textbf{1.000} & \textbf{1.000}
      & \textbf{1.000} & \textbf{1.000} & \textbf{1.000} \\
     \midrule
    \multirow{1}{*}{LLM-based} 
    & PreLog & 0.998 & 0.998 & 0.998 
    & \textbf{0.998} & \textbf{0.998} & \textbf{0.998} 
    & \textbf{0.998} & \textbf{0.998} & \textbf{0.998} 
    & \textbf{0.998} & \textbf{0.998} & \textbf{0.998} 
    & \textbf{0.998} & \textbf{0.998} & \textbf{0.998} \\
    \bottomrule
  \end{tabular}
  }
\end{table*}

\subsection {RQ1: How effective is AnomalyGen in improving model performance across different paradigms?}
To measure how synthetic data volume affects performance, we define an augmentation ratio $R = N_{syn} / N_{real}$ and evaluate at $R \in \{0.001, 0.01, 0.1, 1.0\}$. The baseline uses $R = 0.0$ (no synthetic data). As a concrete example: the HDFS training set has 46,000 sessions, so $R = 0.001$ adds $0.001 \times 46{,}000 = 460$ synthetic sessions. The test set is held fixed across all conditions.

\subsubsection{Results on HDFS}
Table~\ref{tab:evaluation_performance} presents comprehensive results across all 12 models and five augmentation ratios on the HDFS dataset. Adding AnomalyGen data improves performance for every model that is not already at ceiling, with the magnitude of improvement varying by model type and encoding scheme. 

The average F1-score improvement across the seven deep learning models is 2.181\%. The largest gains are for sequence-aware unsupervised models. The Transformer (next\_log) model improves from F1 = 0.818 at baseline to 0.970 at $R = 0.1$, an absolute gain of 15.2 percentage points. This is consistent with our main hypothesis: this model learns a distribution over valid log sequences, and the baseline training data covers too few valid paths to learn a reliable normal distribution. Adding AnomalyGen data exposes the model to many previously unseen but valid sequences, which it can then correctly classify as normal in testing. The LSTM (next\_log) model shows a similar, though smaller, gain: from 0.919 to 0.949 (+3.0\%). Supervised models, which have access to both normal and anomaly labels, benefit more modestly: CNN (sequentials) improves by 1.1\% and LSTM (supervised, sequentials) by about 0.1\%, consistent with the idea that explicit supervision partially compensates for incomplete training data. Notably, the LLM-based PreLog model, despite starting from a high baseline (F1 = 0.989), still improves at $R = 0.01$ to 0.992. Classical ML models (DT, SLFN, KNN) show minor changes within measurement noise which is a result we analyze further in RQ3~\ref{RQ3}.

We also observe that sequence-aware encodings (next\_log, sequentials) consistently benefit more than semantic encodings. For example, Transformer (next\_log) gains 15.2\% while Transformer (semantics) gains only 1.0\%. This reflects a key property of AnomalyGen's generated data: it is structurally correct in terms of execution order, which is exactly the information that sequence-aware encodings capture. Semantic encodings aggregate log content and discard ordering information, so they benefit less from the structural validity of the generated sequences.

\subsubsection{Results on Zookeeper}
Table~\ref{tab:evaluation_performance_zookeeper} presents results on the Zookeeper dataset, further validating AnomalyGen's generalizability across different systems. The average improvement is 1.692\% with a maximum of 13.0\% for LSTM (next\_log) and 8.2\% for Transformer (next\_log, from 0.610 to 0.660). The Zookeeper results confirm that AnomalyGen's benefits are not dataset-specific. LSTM (next\_log) improves from 0.624 to 0.705 (+13.0\%), and Transformer (next\_log) improves from 0.610 to 0.660 (+8.2\%). For models that already perform well (DT, KNN, SLFN achieving 1.000 F1-scores, and PreLog achieving 0.998), AnomalyGen maintains high performance without degradation, demonstrating the high fidelity of our augmented data. 

Comparing the two systems, HDFS shows larger absolute improvements than Zookeeper. HDFS is a more complex distributed system with a larger codebase and more diverse execution paths; the baseline dataset covers only 0.99\% of its source code templates, compared to 8.02\% for Zookeeper. The more severe the initial coverage gap, the larger the benefit from code-guided augmentation.

\begin{center}
\fbox{\begin{minipage}{0.9\linewidth}
\textbf{Answer to RQ1:} AnomalyGen consistently improves performance across all 12 evaluated models and both systems. On HDFS, the average F1-score improvement for deep learning models is 2.181\%, and on Zookeeper, the average improvement is 1.692\%.
\end{minipage}}
\end{center}

\begin{table}[h]
\centering
\caption{Ablation experiments on the static analysis (Phase I), CoT reasoning (Phase II), and knowledge-driven annotation (Phase III). Bold indicates the best performance.}
\label{tab:ablation}
\begin{tabular*}{\textwidth}{@{\extracolsep{\fill}}l|ccc|ccc|ccc|ccc@{}}
\toprule
 & \multicolumn{3}{c|}{AnomalyGen} 
 & \multicolumn{3}{c|}{w/o analysis} 
 & \multicolumn{3}{c|}{w/o cot} 
 & \multicolumn{3}{c}{w/o label} \\
\cmidrule(lr){2-4} 
\cmidrule(lr){5-7} 
\cmidrule(lr){8-10}
\cmidrule(lr){11-13}
Model & F1 & RC & PC & F1 & RC & PC & F1 & RC & PC & F1 & RC & PC \\
\midrule
transformer-next\_log 
& \textbf{0.921} & \textbf{0.984} & \textbf{0.866}
& 0.834 & 0.733 & 0.966
& 0.814 & 0.731 & 0.919
& 0.854 & 0.793 & 0.927 \\

transformer-semantic
& \textbf{0.725} & \textbf{0.632} & \textbf{0.850}
& 0.697 & 0.561 & 0.920
& 0.709 & 0.560 & 0.964
& 0.704 & 0.579 & 0.899 \\

cnn-semantic
& \textbf{0.971} & \textbf{0.973} & \textbf{0.969}
& 0.968 & 0.989 & 0.948
& 0.964 & 0.986 & 0.943
& 0.972 & 0.985 & 0.959 \\

cnn-sequentials
& \textbf{0.971} & \textbf{0.995} & \textbf{0.949}
& 0.965 & 0.995 & 0.936
& 0.965 & 0.995 & 0.935
& 0.961 & 0.995 & 0.928 \\

lstm-next\_log
& \textbf{0.925} & \textbf{0.891} & \textbf{0.963}
& 0.901 & 0.848 & 0.961
& 0.910 & 0.926 & 0.894
& 0.909 & 0.930 & 0.889 \\

lstm-sequentials
& \textbf{0.968} & \textbf{0.995} & \textbf{0.943}
& 0.901 & 0.848 & 0.961
& 0.960 & 0.996 & 0.926
& 0.958 & 0.995 & 0.924 \\

lstm-semantic-next\_log
& \textbf{0.760} & \textbf{0.623} & \textbf{0.975}
& 0.697 & 0.561 & 0.920
& 0.717 & 0.598 & 0.895
& 0.718 & 0.622 & 0.850 \\

lstm-semantic
& \textbf{0.968} & \textbf{0.988} & \textbf{0.948}
& 0.967 & 0.988 & 0.948
& 0.962 & 0.995 & 0.931
& 0.968 & 0.988 & 0.949 \\

llm-based
& \textbf{0.990} & \textbf{0.991} & \textbf{0.990}
& 0.955 & 0.941 & 0.970
& 0.955 & 0.941 & 0.970
& 0.955 & 0.942 & 0.970 \\

\bottomrule
\end{tabular*}
\end{table}

\subsection{RQ2: What is the contribution of each core component of AnomalyGen?}
To systematically assess the contribution of each key component in AnomalyGen's three-stage pipeline, we conduct a comprehensive ablation study (shown in Table~\ref{tab:ablation}) with three carefully designed variants as below. All variants use augmentation ratio 0.001 on HDFS, and we evaluate nine deep learning and LLM-based models (classical ML models are excluded because they are insensitive to augmentation, as discussed in RQ3~\ref{RQ3}). The three variants are: (1) \textit{w/o analysis}, which omits the LCFG construction in Phase I and uses the raw call graph instead; (2) \textit{w/o cot}, which replaces the CoT verification in Phase II with direct LLM generation without step-by-step reasoning; (3) \textit{w/o label}, which replaces knowledge-driven labeling with a simple severity-level rule (any sequence containing an \texttt{[ERROR]} log is labeled anomalous).

Understanding each component's importance is critical for knowing what makes code-guided augmentation effective and practical guidance on deployment priorities when computational resources are limited.

\subsubsection{CoT Verification: The Most Critical Component}
Removing CoT verification causes the largest performance drop across nearly all models. The Transformer (next\_log) drops from F1 = 0.921 to 0.814 ($-10.7$ points); the LSTM (next\_log) drops from 0.925 to 0.910 ($-1.5$ points). Examining the generated data explains why. Without CoT reasoning, the generated data has a severely skewed label distribution: only 30 normal sessions vs.\ 69 anomalous ones (30\% normal rate), compared to AnomalyGen's balanced 57 normal vs.\ 49 anomalous (54\% normal rate). When AnomalyGen cannot distinguish logically valid path combinations from invalid ones, it generates sequences that look structurally connected but represent impossible executions (e.g., error handling code appearing without any preceding error condition). These sequences are labeled anomalous, giving an inflated anomaly proportion in training. This skew is particularly damaging for unsupervised models, which learn a normality model and are sensitive to the ratio of normal to anomalous sequences. Supervised models show smaller degradation (CNN: $-0.6$ to $-0.7$ points) because their explicit labels provide a corrective signal.

\subsubsection{LCFG Construction: Capturing System Complexity}
Omitting LCFG construction (w/o analysis) leads to substantial degradation, particularly for unsupervised models. Transformer (next\_log) falls from 0.921 to 0.834 ($-8.7$ points, $-9.4$\% relative), while LSTM (next\_log) decreases from 0.925 to 0.901 ($-2.4$ points). The LCFG captures fine-grained control flow within methods, including conditional branches, loops, and exception handlers. Without this fine-grained analysis, the method misses \textit{rare but valid execution paths}—those edge cases occurring under specific runtime conditions (e.g., "allocation fails → retry → succeeds"). These rare paths are exactly what sparse baseline data lacks and what AnomalyGen is designed to provide. Supervised models are less affected (CNN: $-0.1$ to $-0.6$ points), as they can learn decision boundaries from labeled data even when some edge cases are missing. Interestingly, PreLog also shows a large degradation ($-3.5$ points), suggesting that LLMs benefit from seeing diverse execution patterns as examples for in-context learning. LCFG construction (Phase I) ensures the space of possible paths is sufficiently rich and diverse. CoT verification (Phase II) ensures paths sampled from this space are logically coherent. Both are necessary: LCFG without CoT produces diverse but incoherent paths; CoT without LCFG produces coherent but insufficiently diverse paths.

\subsubsection{Knowledge-Driven Labeling: Quality Over Heuristics}
Replacing knowledge-driven labeling with simple severity-level rules causes Transformer (next\_log) to drop from 0.921 to 0.854 ($-6.7$ points) and LSTM (next\_log) from 0.925 to 0.909 ($-1.6$ points). The impact on unsupervised models is indirect: the simple labeling rule over-identifies paths as anomalous (any \texttt{[ERROR]} message triggers an anomaly label, including normal retry paths that log errors before recovering). This reduces the diversity of normal training sequences, recreating the incomplete coverage problem. Supervised models are nearly unaffected (CNN: 0 to $-1.0$ points) because they can fall back on the original labeled data.

Each component makes different contributions. CoT verification is the most critical: removing it causes $-10.7$ points for the Transformer and produces a severely skewed normal/anomaly ratio in training data. LCFG construction is the second most important: removing it causes $-8.7$ points by preventing the generation of rare but valid execution paths. Knowledge-driven labeling affects even unsupervised models indirectly ($-6.7$ points) by reducing normal path diversity. These ablation results have important implications for future data augmentation methods in log anomaly detection: Methods that generate synthetic logs without ensuring logical coherence (e.g., through template-based randomization or simple LLM prompting) will likely introduce harmful noise that degrades model performance, particularly for unsupervised learners. 


\begin{center}
\fbox{\begin{minipage}{0.9\linewidth}
\textbf{Answer to RQ2:} All three AnomalyGen components contribute to performance. CoT verification is the most critical: removing it causes $-10.7$ points for the Transformer, and LCFG construction is the second most important: removing it causes $-8.7$ points.
\end{minipage}}
\end{center}


\subsection{RQ3: How do different experimental settings impact AnomalyGen's effectiveness?}
\label{RQ3}
The response to AnomalyGen augmentation varies substantially across model paradigms, and understanding these differences provides practical guidance for deployment.

\begin{figure}[h]
\centering
\includegraphics[width=0.7\linewidth]{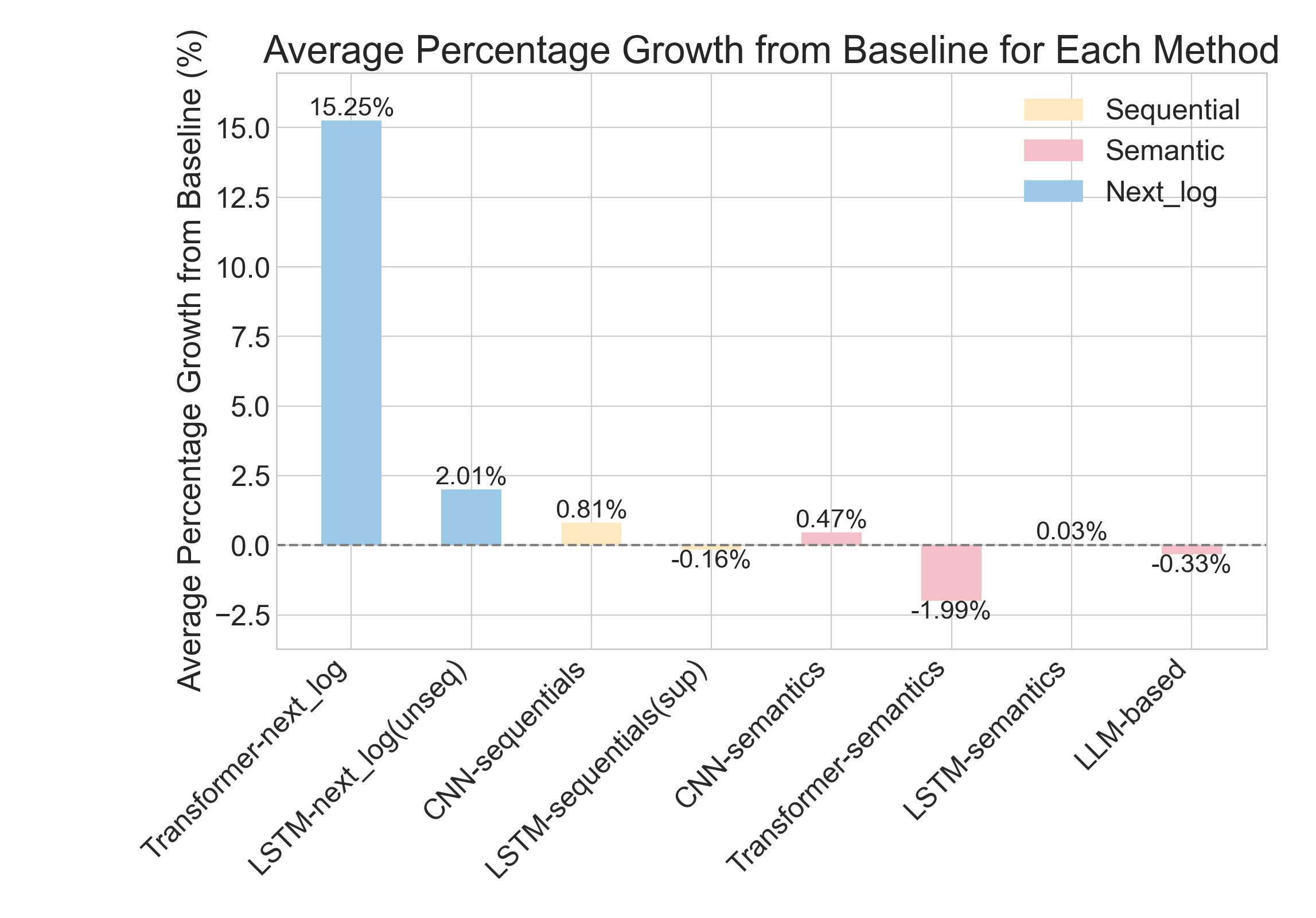}
\caption{Average F1-score improvement under different encoding schemes across model paradigms.}
\label{fig:encoding_comparison}
\end{figure}

\subsubsection{Unsupervised Models.}
\begin{figure}
    \centering
    \begin{subfigure}{0.46\textwidth}
        \centering
        \includegraphics[width=\textwidth]{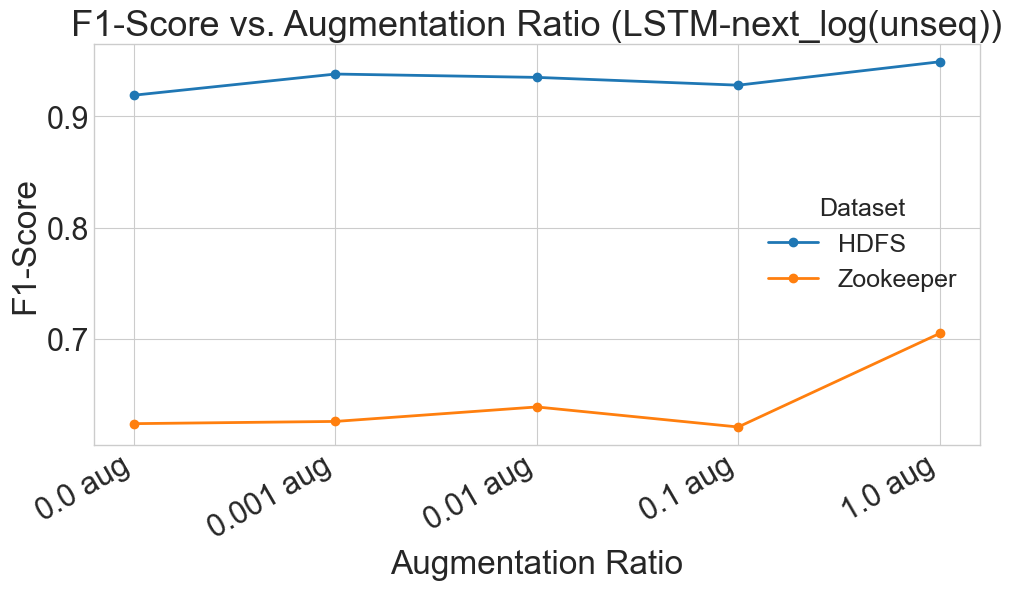}
        \vspace{-10pt}
    \end{subfigure}
    \begin{subfigure}{0.46\textwidth}
        \centering
        \includegraphics[width=\textwidth]{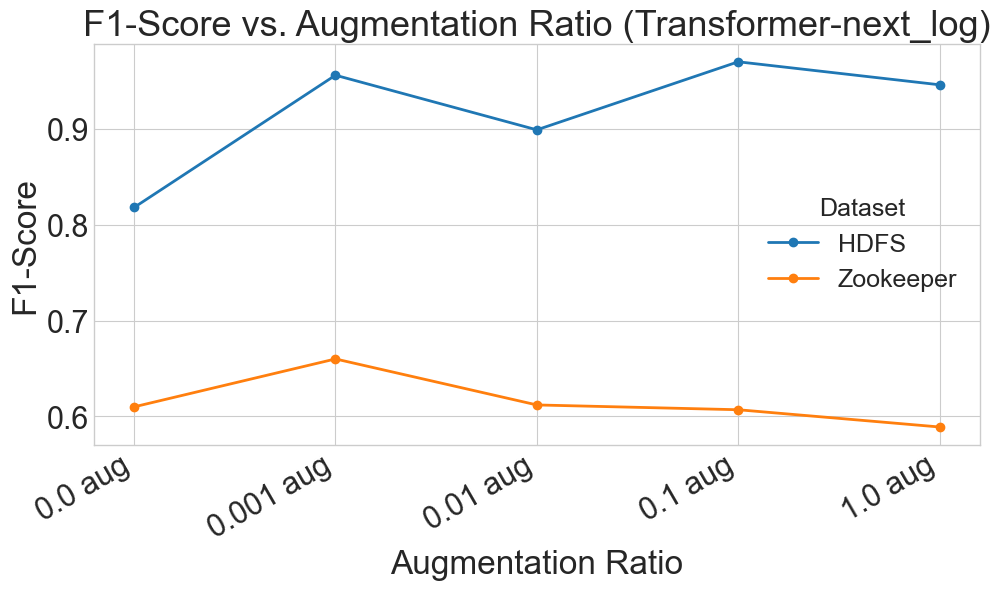}
        \vspace{-10pt}
    \end{subfigure}
    \caption{UnSupervised Deep Learning Performance Under Different Augmentation Ratio}
    \vspace{-15pt}
    \label{fig:RQ2-file Modification in Unsupervised}
\end{figure}

Figure~\ref{fig:RQ2-file Modification in Unsupervised} shows that unsupervised models exhibit a non-monotonic response to augmentation ratio. On HDFS, the Transformer (next\_log) improves sharply from 0.818 at baseline to 0.956 at $R = 0.001$ (+16.9\%), peaks at 0.970 at $R = 0.1$ (+18.6\%), then declines slightly to 0.946 at $R = 1.0$ (+15.6\%). This pattern can be explained by the trade-off between coverage enrichment and distribution shift. At low ratios, the synthetic data broadens the model's training distribution without altering its statistical properties significantly. At high ratios ($R = 1.0$), the synthetic data's anomaly rate (~50\%) diverges from the test set's anomaly rate (~15\%), shifting the training distribution enough to confuse the model. The practical implication is that a moderate augmentation ratio (0.01--0.1) is optimal for unsupervised models, and this ratio should be tuned for each system and architecture. 

Additionally, the encoding scheme has a dramatic impact on unsupervised models. Figure~\ref{fig:encoding_comparison} shows that Transformer (next\_log) achieves +15.2\% improvement, while Transformer (semantics) gains only +1.0\%—a 14.2 percentage point gap. This stark difference reveals AnomalyGen's core value proposition: it excels at generating structurally valid, temporally coherent execution paths. Sequence-aware encoding patterns (next\_log, sequential) preserve the temporal ordering and execution flow information that AnomalyGen generates. These patterns allow models to learn realistic execution patterns—e.g., ``log event A should be followed by log event B under normal conditions.'' In contrast, semantic encoding aggregates log content and discard temporal ordering, losing the critical structural information that makes AnomalyGen effective.

\subsubsection{Supervised Models: Stable Performance with Balanced Encoding Benefits}
\begin{figure}
    \centering
    \begin{subfigure}{0.46\textwidth}
        \centering
        \includegraphics[width=\textwidth]{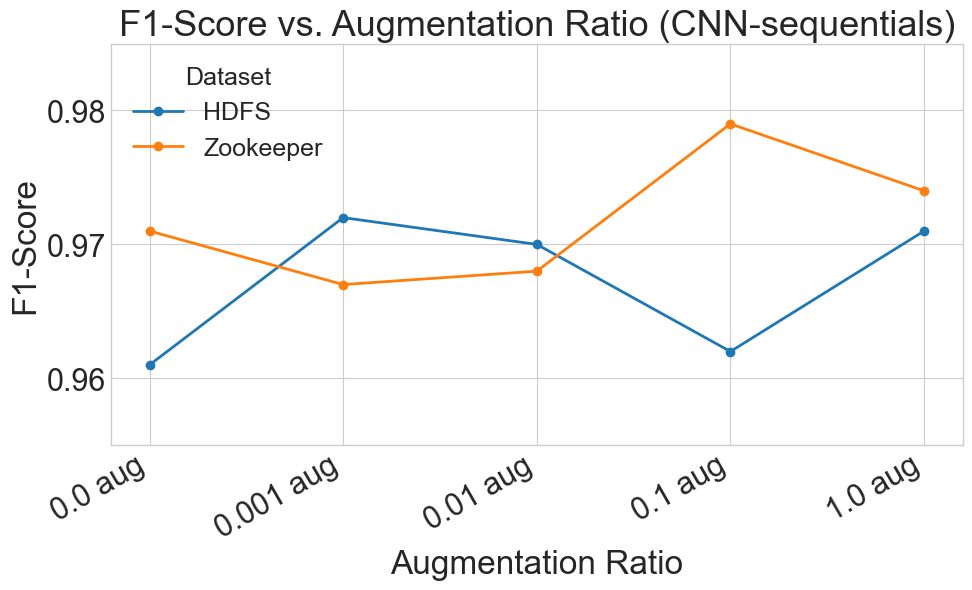}
        \vspace{-10pt}
    \end{subfigure}
    \begin{subfigure}{0.46\textwidth}
        \centering
        \includegraphics[width=\textwidth]{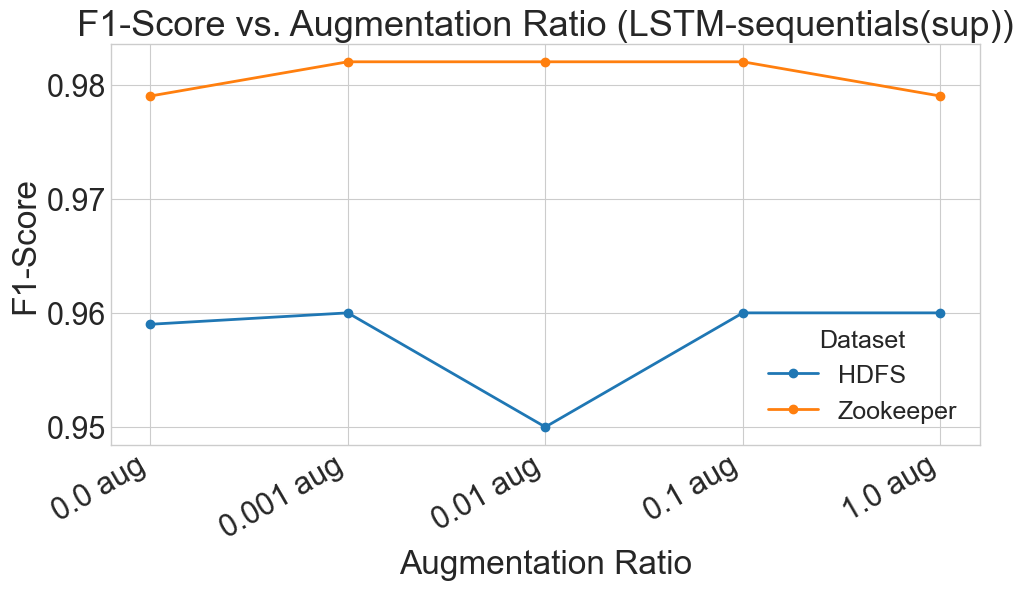}
        \vspace{-10pt}
    \end{subfigure}
    \caption{Supervised Deep Learning Performance Under Different Augmentation Ratio}
    \vspace{-15pt}
    \label{fig:RQ2-file Modification in Supervised}
\end{figure}

Figure~\ref{fig:RQ2-file Modification in Supervised} shows that supervised models are much less sensitive to the augmentation ratio. CNN (sequentials) peaks at $R = 0.001$ (+1.1\%) and remains stable across higher ratios. LSTM (sequentials) shows similar behavior, with only minor fluctuations across different ratios. This stability comes from the explicit labels: even if the augmented training distribution differs from the test distribution, the labels provide a direct training signal that prevents the model from being misled by the distribution shift. 

Unlike unsupervised models, supervised models benefit from AnomalyGen across both encoding types, though with different magnitudes. CNN (sequentials) improves by +1.1\%, while CNN (semantics) gains +0.7\%—a 0.4 percentage point gap, much smaller than the 14.2-point gap observed for unsupervised Transformer.
The explicit labels allow supervised models to leverage both dimensions of AnomalyGen's enhancements: (1) structurally correct execution paths benefit sequential encodings by providing realistic temporal patterns, and (2) realistic parameter values and log messages generated by LLM-based reasoning benefit semantic encodings by enriching the vocabulary and semantic patterns. The label guidance makes the model less critically dependent on a single information source. For supervised models, augmented data acts as a form of regularization. It introduces novel, structurally valid edge cases that are not present in the original training set, helping the model refine its decision boundary and reduce overfitting. This explains why even small amounts of augmentation (ratio 0.001) can improve performance, and why performance remains stable at higher ratios—the model can effectively use labels to separate signal from noise.

\subsubsection{LLM-based Models: Few-Shot Learning with Low-Ratio Optimization}
PreLog peaks at $R = 0.01$ on HDFS (+0.3\%) but degrades at higher ratios (0.977 at $R = 0.1$). This pattern differs from both supervised and unsupervised models and reflects the in-context learning mechanism: a small number of high-quality synthetic examples refines the model's decision boundary, while too many examples may shift the statistical distribution in the prompt context. On Zookeeper, PreLog is robust across all ratios (constant 0.998), suggesting that for simpler tasks, LLMs are less sensitive to augmentation volume. 

\subsubsection{Classical ML Models: Foundational Baselines with Complete Insensitivity}
\begin{figure}
    \centering
    \begin{subfigure}{0.46\textwidth}
        \centering
        \includegraphics[width=\textwidth]{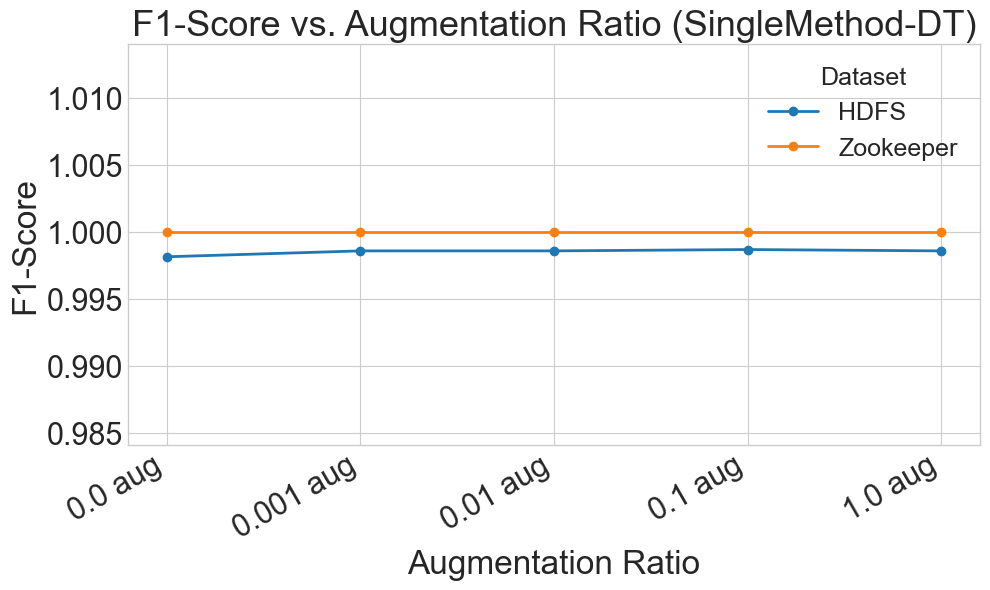}
        \vspace{-10pt}
    \end{subfigure}
    \begin{subfigure}{0.46\textwidth}
        \centering
        \includegraphics[width=\textwidth]{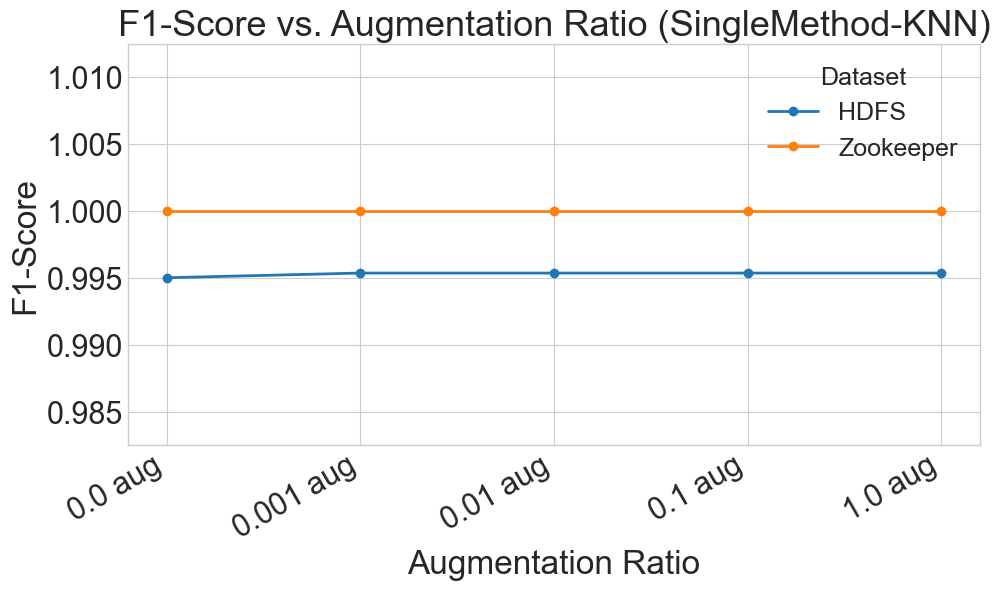}
        \vspace{-10pt}
    \end{subfigure}
    \caption{Machine Learning Performance Under Different Augmentation Ratio}
    \vspace{-15pt}
    \label{fig:RQ2-file Modification in ML}
\end{figure}

Figure~\ref{fig:RQ2-file Modification in ML} shows that on Zookeeper, DT, KNN, and SLFN maintain perfect 1.000 F1-scores across all ratios (0.0, 0.001, 0.01, 0.1, 1.0). On HDFS, minor fluctuations in the third or fourth decimal place are within measurement noise (e.g., DT: 0.99815 → 0.99869 → 0.99859). These models rely on simple, frequency-based features, essentially counting how many times each log template appears in a sequence. AnomalyGen generates sequences with rich structural information (execution paths, temporal relationships), but these nuances are invisible to frequency-based feature vectors. As long as the overall template frequency distribution remains statistically consistent, classical models cannot distinguish original from augmented data. The insensitivity also reflects that classical models have reached their performance ceiling on these datasets, as noted by prior work~\cite{yu2024deep}. They cannot leverage the additional structural information that AnomalyGen provides because their feature representations are too simplistic.

\begin{center}
\fbox{\begin{minipage}{0.9\linewidth}
\textbf{Answer to RQ3:} Model paradigm is the primary determinant of how a model responds to AnomalyGen augmentation. Unsupervised models benefit most but require careful ratio tuning, whereas classical ML models are insensitive to augmentation.
\end{minipage}}
\end{center}

\subsection{RQ4: What is the computational overhead of AnomalyGen in practical scenarios?}

The practical utility of a data augmentation framework is determined not only by its effectiveness but also by its generation cost. Prohibitive overhead can hinder its adoption in real-world engineering contexts. This research question aims to quantify the computational overhead of AnomalyGen and conduct a cost-benefit analysis against traditional data acquisition methods (e.g., manual labeling, live system deployment). By decomposing the cost structure of AnomalyGen, we evaluate its feasibility for practical applications. 
The generation cost of AnomalyGen is composed of two primary components: a one-time static analysis cost and an on-demand sequence generation cost as shown in Table~\ref{tab:overhead}.

\begin{table}[h]
\small
\centering
\caption{Overhead Statistics of Components in Each Stage} 
\label{tab:overhead}
\begin{tabular}{@{}lll@{}}
\toprule
\textbf{Stage} & \textbf{Component} & \textbf{Overhead} \\
\midrule
\multirow{2}{*}{One-Time Static Analysis Cost} & Compile & \textasciitilde 9 minutes \\
& Call Graph Generation and Pruning & \textasciitilde 1.5 minutes \\
\midrule
\multirow{2}{*}{On-Demand Sequence Generation Cost} & Total Time Cost (Single Processor) & \textasciitilde 373 seconds / entry \\
& Number of LLM API Calls & \textasciitilde 30 times / entry  \\
\bottomrule
\end{tabular}
\end{table}
The one-time static analysis cost is minimal, requiring approximately 9 minutes for compiling Hadoop codebase and approximately 1.5 minutes (69.08s + 23.70s) for call graph generation and pruning. This initial investment is amortized, as the resulting pruned graph can be reused indefinitely for all subsequent generation tasks.

The on-demand generation cost is primarily driven by LLMs interactions. For precise measurement, we randomly selected 11 entry points and performed single-process generation with a depth of 3. During this process, we observed that the generation of an average sequence requires around 30 interactions with the LLM API. While a single-process baseline takes about 373 seconds per entry (including fine-grained analysis and LLM calls to generate multiple sequences), this task is embarrassingly parallel. On our 32-core server, the generation process can be massively accelerated by leveraging multi-threading, with the primary bottleneck shifting from local computation to API network latency and rate limits. 
This scalability is crucial for practical applications. 

Compared to traditional data collection, AnomalyGen is substantially faster. Prior work reports spending 26 days collecting 200,000 logs for Zookeeper~\cite{he2020loghub} and 38.7 hours to collect 30 log templates from HDFS~\cite{zhu2023loghub}. Manual labeling of collected logs requires additional expert time. AnomalyGen replaces this process with a largely automated pipeline whose main cost is LLM API usage.

\begin{center}
\fbox{\begin{minipage}{0.9\linewidth}
\textbf{Answer to RQ4:} AnomalyGen is highly efficient and scalable. It has a low, one-time setup cost of under 15 minutes and a parallelized on-demand generation cost, confirming its practical utility for industrial applications.
\end{minipage}}
\end{center}


\section{Threats to Validity}
We acknowledge several potential threats to the validity of our study.

\textbf{LLM reasoning uncertainty.}  
AnomalyGen relies on the reasoning capabilities of LLMs during Phase II. We recognize that LLMs are stochastic and not logically infallible; there remains a residual risk of hallucinating unreachable paths, incorrectly resolving dynamic calls, or producing semantically plausible but structurally incorrect sequences. To mitigate this, we employed well-established LLMs (GPT-4o and DeepSeek-V3) and utilized CoT prompting~\cite{wei2023chainofthoughtpromptingelicitsreasoning}, a technique proven to enhance reasoning accuracy. Additionally, we set temperature to 0 to eliminate randomness and ensure reproducibility. Future work could further reduce this uncertainty by employing ensemble methods or integrating formal verification checkers alongside LLM reasoning.

\textbf{Completeness of anomaly labels.} 
Phase III labeling rules are grounded in common engineering practice and cover the most frequent anomaly patterns, but they will not capture every anomaly type, particularly zero-day faults or failures expressed through application-specific semantics. The rule base is designed to be extensible: as new fault patterns are identified, new rules can be added. Our manual validation on 247 sessions (95.28--95.74\% accuracy) provides empirical evidence of the rule quality.

\textbf{Source code availability.} We acknowledge that AnomalyGen requires access to source code, which may be perceived as a limitation for analyzing closed-source third-party software. However, we explicitly position AnomalyGen as a white-box verification tool for DevOps and Site Reliability Engineering (SRE) teams, rather than a passive black-box monitor. In modern CI/CD pipelines, developers possess full access to their codebase. AnomalyGen is designed to be integrated into the build process, automatically generating log sequences to train monitoring models before deployment. Regarding third-party libraries (e.g., JAR files), this limitation is not fundamental; future iterations can integrate lightweight bytecode analysis tools (such as Soot or ASM) to extract CFGs from compiled binaries without requiring raw source code.

\textbf{Language generalizability.} Our evaluation covers Java systems (HDFS and Zookeeper). The three-phase methodology is language-agnostic in principle, but the Phase I implementation is Java-specific (using JavaParser). Adapting AnomalyGen to other languages (C++, Python, Go) requires replacing the front-end analysis tools with language-appropriate alternatives (e.g., Clang for C/C++ or Python's AST module), which is an engineering task rather than a conceptual one and we leave as a direction for future research.

\section{Conclusion}
We presented AnomalyGen, a framework for augmenting log anomaly detection data with sequences synthesized from source code. 
The key observation motivating our work is that popular benchmarks cover less than 10\% of the execution behaviors defined in source code; as a result, detection models trained on these benchmarks frequently misclassify unseen but valid execution paths as anomalies. Simpler augmentation strategies (such as perturbation and random resampling) fail to address this gap and often degrade performance. AnomalyGen addresses this by using static analysis to define the space of valid execution paths and LLM-based reasoning to verify their logical consistency and generate plausible parameters.

Our experiments across 12 detection models on HDFS and Zookeeper confirm that increasing training data coverage consistently improves detection performance. This success stems from its ability to generate logically coherent execution paths that respect program structure. Our analysis revealed that unsupervised, sequence-aware models are the primary beneficiaries, especially when paired with a moderate augmentation ratio. An ablation study further confirmed that CoT verification is the most critical component for ensuring data quality. Finally, we showed that AnomalyGen is highly practical, with a low setup cost and a scalable generation process, making it orders of magnitude more efficient than traditional methods. This work underscores that for data augmentation to be effective, it must be grounded in the underlying logic of the source program.

Future work will extend AnomalyGen to additional programming languages and explore the use of symbolic execution to handle complex data-dependent paths that are difficult for LLM reasoning alone.

\section{Data Availability}
All code, datasets, and model settings (logs) for AnomalyGen are available at \url{https://github.com/2ira/AnomalyGen-main} for replication and follow-up research.

\bibliographystyle{ACM-Reference-Format}
\bibliography{sample-base}

\end{document}